\documentclass[aps,jcp,twocolumn,superscriptaddress,showpacs]{revtex4-2}
\usepackage{natbib}
\usepackage[english]{babel}
\usepackage[dvips]{graphics}
\usepackage{graphicx,epsfig}
\usepackage{amsmath}
\usepackage{color}
\usepackage{multirow}
\usepackage[normalem]{ulem}
\usepackage{booktabs}
\usepackage{amsfonts} 
\usepackage{multirow}
\usepackage{lineno}
\newcommand{\SNu}{SN$_\textnormal{u}$}

\begin{document}
\title{Identification of the invariant manifolds of
the LiCN molecule using Lagrangian descriptors}

\author{F. Revuelta}
\affiliation{Grupo de Sistemas Complejos, 
    Escuela T\'ecnica Superior de Ingenier\'ia Agron\'omica, Alimentaria y de Biosistemas,
    Universidad Polit\'ecnica de Madrid,
    Avda.\ Puerta de Hierro 2-4, 28040 Madrid, Spain.} 
\author{R. M. Benito}
\affiliation{Grupo de Sistemas Complejos, 
    Escuela T\'ecnica Superior de Ingenier\'ia Agron\'omica, Alimentaria y de Biosistemas,
    Universidad Polit\'ecnica de Madrid,
    Avda.\ Puerta de Hierro 2-4, 28040 Madrid, Spain.} 
\author{F. Borondo}
\affiliation{Instituto de Ciencias Matem\'aticas (ICMAT), 
    Cantoblanco, 28049  Madrid, Spain.} 
\affiliation{Departamento de Qu\'imica, 
    Universidad Aut\'onoma de Madrid, Cantoblanco, 28049 Madrid, Spain.}

\date{\today}

\begin{abstract}
    In this paper, we apply 
Lagrangian descriptors to study 
    the invariant manifolds
    that emerge from the top of 
    two  barriers existing in the LiCN$\rightleftharpoons$LiNC 
    isomerization reaction.
    We demonstrate that the integration times must be large enough
    compared with the characteristic stability exponents of the
     periodic orbit under study.
    The invariant manifolds  manifest as singularities in the
    Lagrangian descriptors.
%
     Furthermore,  we develop an 
     equivalent potential energy surface with 2 degrees of freedom,
    which reproduces with a great accuracy
    previous results [\emph{Phys. Rev. E} \textbf{99}, 032221 (2019)].
    This 
   surface allows the use of an adiabatic approximation to
    develop a
    more simplified 
    potential energy  with solely 1 degree of freedom.
    The reduced dimensional model is 
    still able to qualitatively describe the results observed 
    with the original 2-degrees-of-freedom potential energy landscape.
    Likewise, it
    is also used to study in a more simple manner
    the influence on the Lagrangian descriptors
    of a bifurcation,
    where some of the previous invariant manifolds  emerge,
    even before it takes place.
\end{abstract}

\maketitle

\section{Introduction} \label{sec:intro}

Molecular systems usually exhibit a very rich and intricate dynamics,
even in small molecules formed solely by a few atoms~\cite{Levine75},
due to nonlinear interactions~\cite{Karmakar20}.
The existence of conical intersections~\cite{Worth04, Polli10} between 
Born-Oppenheimer potential energy surfaces (PES) adds additional 
complexity to the classical characterization of these systems, 
as molecules may undergo electronic transitions in their neighborhoods
\cite{Jiang12}.
Likewise, the combination of these effects easily opens effective 
routes to chaos, 
making  the analysis of the corresponding dynamics more
complex, in particular when bifurcations take place
and substantially modify the structure of the phase space.

Suitable tools
to cope with the previous problems 
can be borrowed from dynamical systems theory
\cite{LL10}, which sets up the molecular phase space as the
proper arena for a dynamical analysis.
Accordingly, for low excitation energies the motion of the nuclei takes place
in the vicinity of the lowest equilibrium point of the PES, where the
harmonic approximation is valid and the motions
are well characterized by 
the normal modes~\cite{Levine75}.
The structure of the corresponding phase space is mostly regular,
with all motions organized in invariant tori.
As the excitation energy increases, the anharmonicities 
and the coupling among the different modes open
 the door to irregular motion and effective
intramolecular vibrational relaxation~\cite{Fujisaki05}.
From a nonlinear dynamics perspective,
these phenomena can be partially explained through
the Kolmogorov-Arnold-Moser 
theorem~\cite{Anold06}, which
dictates that the perturbation associated with the energy growth destroys
some of the previous tori.
In addition, 
the Poincar\'e-Birkhoff 
theorem~\cite{Birkhoff13}
locally controls the vicinity of the regions
where resonances among modes are important~\cite{Jiang12}.
%
Eventually, the vibrational energy can 
become large enough
to 
classically overcome 
energetic barriers (typically saddles) in the PES;
this gives rise to
chemical reactivity~\cite{Marcus92}.

The study of chemical reactions from a dynamical perspective 
dates back to Marcelin's~\cite{Marcelin15} and Wigner's~\cite{Wigner38} pioneering works
on transition state theory~\cite{Truhlar96},
which are also applicable to other physical processes in which the system can be partitioned 
into different regions~\cite{Jaffe00, Jaffe02, Uzer02, deOliveira02}.
The key point is the study of the dynamics at the top of the barrier separating 
two such parts (usually referred to as the \textit{reactants} and the \textit{products}),
where some relevant geometrical structures can be identified,
i.e., a \textit{normally hyperbolic invariant manifold} 
\cite{Uzer02}
and its invariant manifolds~\cite{Gorban05}, which 
locally
determine the 
classical reactive dynamics.
Moreover, 
at the quantum level,
tunneling and interference might be  
important~\cite{Miller98, Althorpe13, Jang16, Hele16}.
Early examples of 
classical structures in two-dimensional problems were envisioned, 
for example, in Ref.~\onlinecite{Pechukas77}.
These concepts have also been generalized to the case of noisy driven systems,
\cite{Bartsch05,Kawai07,Bartsch12, Revuelta12, Craven15, Bartsch19}, 
using a chaotic trajectory that jiggles around the barrier top stochastically.
Based on these theories, many studies to understand chemical reactivity 
have been conducted considering only the barrier top on the PES, 
since this is the most important dynamical bottleneck.
However, other 
phase-space
\textit{dynamical} barriers 
\cite{Davis81,Borondo95,Borondo96,Zembekov97} may exist interfering with
the motion in this region, 
and modulating, for example, the corresponding
time scales~\cite{Muller12}.
The correct identification of the geometrical structures, i.\,e.,
the invariant manifolds, in this other situation
is similarly of great importance.

The aim of this paper is 
the identification of the invariant manifolds
that emerge from the top of 
two  barriers existing in the LiCN$\rightleftharpoons$LiNC isomerization
reaction,
paying special attention to those of dynamical origin.
These manifolds act as true geometrical separatrices for the system.
For this purpose, 
we  use the Lagrangian descriptors (LDs)~\cite{Madrid09, Lopesino15}, 
a recently developed tool that allows the study of the classical flow of 
dynamical systems in a very simple and effective way.

First, 
we introduce an alternative PES
with 2 degrees of freedom (dof) that is
formed by Morse potentials.
We demonstrate that the results obtained using this PES
are in excellent agreement
with those yielded by the original \emph{ab initio} one,
even when a constant moment of inertia is considered
(see discussion in Sec.~\ref{sec:Vequiv2dofs}).

Second,
in order to avoid the complicated picture derived from
excessive homoclinic and heteroclinic intersections,
in the previous PESs,
which have 2 dof
we make use of the adiabatic approximation
in order to obtain 
an equivalent 
1-dof model,
something that is feasible
due to the great performance of 
PES formed by Morse oscillators.
The reduced dimensional model
is nevertheless  able 
to capture the phenomenology that takes place in the vicinity of the barriers
existing in the system
and, as a consequence,
provides an adequate characterization
of its dynamics (see Sec.~\ref{sec:1dof}).

This paper is organized as follows.
After the Introduction, we present in Sec.~\ref{sec:system} 
the system under study, 
highlighting the main dynamical peculiarities
and four periodic orbits (POs) that are relevant for our work.
Section~\ref{sec:LD} briefly describes
the method used to
unveil the structures existing in the vibrational phase space of our system.
Next, we present in Sec.~\ref{sec:results} the main results of our work,
paying special attention to the geometry surrounding the dynamical
barrier, along with the corresponding discussion.
Finally, we conclude the paper
in Sec.~\ref{sec:summary} with the Summary and Outlook.

\section{System and classical trajectories} \label{sec:system}
In this section
we report the main properties of the isomerizing reaction under study.
To begin,
we present in Sec.~\ref{sec:H} the Hamiltonian function 
that models the system.
Next, 
in Sec.~\ref{sec:2dof_Vequiv},
we describe an alternative PES
formed by a collection of Morse oscillators.
obtained by making use of an adiabatic approximation.
Finally,
Sec.~\ref{sec:dyn} is devoted to a brief description of
the vibrational dynamics of the studied molecule.

\subsection{Hamiltonian model with 2 degrees of freedom}
\label{sec:H}
\begin{figure}
\includegraphics[width=0.85 \columnwidth]{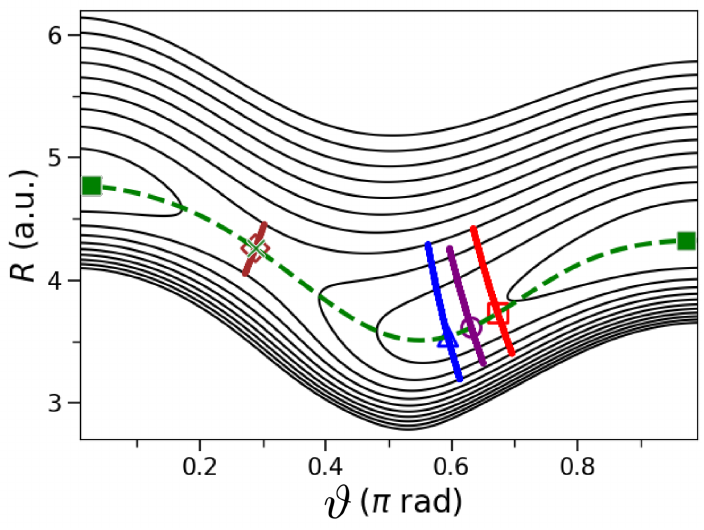}
\caption{\emph{Ab initio} potential energy surface for LiCN molecule
with 2 degrees of freedom.
It has two minima
(green squares) corresponding to the (stable) linear configurations
Li-C$\equiv$N ($\vartheta = 0$\,rad) and C$\equiv$N-Li 
($\vartheta = \pi$\,rad), which are connected by the minimum energy path
(dashed green line) for the isomerization reaction,
which passes through a rank-1 saddle (green cross contained in the brown diamond).
Four periodic orbits relevant to our work are also shown.
The remaining symbols highlight the intersection of the previous
trajectories with the minimum energy path; 
this defines the Poincar\'e
surface of section,
where initial conditions are taken (see text for details).
}
\label{fig:1}
\end{figure}

The system under study is the rotationless ($J=0$) triatomic LiCN molecule.
Here, the motion associated with the triple bond in the C$\equiv$N fragment 
has a very high frequency, and
consequently 
decouples
from the rest 
of the molecular modes.
Accordingly, the 
CN stretch dof
can be assumed to remain constant at
its equilibrium value,~$r_{\rm eq}= 2.186$\,a.u.
The vibrations of the LiCN can then be very well  described 
with the following 2-dof
Hamiltonian:
\begin{equation}  \label{eq:H_2dof}
  \mathcal{ H }_2 = \frac{P_R^2}{2 \mu_1} 
  +  \frac{P_\vartheta^2}{2} \left( \frac{1}{\mu_1 R^2} 
  + \frac{1}{\mu_2 r_{\rm eq}^2} \right)
  + V_{\rm{a.i.}} (R, \vartheta),
\end{equation}
where the distance~$R$ and the angle~$\vartheta$ determine the position 
of the 
Li atom with respect to the center of mass of C-N,
$\mu_1 = m_{\rm Li} (m_{\rm C} + m_{\rm N}) /
( m_{\rm Li} + m_{\rm C} + m_{\rm N})$,
and~$\mu_2=(m_{\rm C}m_{\rm N})/(m_{\rm C}+m_{\rm N})$ are reduced masses,
and~$V_{\rm{a.i.}}   (R, \vartheta)$ is the 
\emph{ab initio}
PES~\cite{Essers82},
which is shown in Fig.~\ref{fig:1} in the form of a contours plot.
As can be seen, this PES has two stable minima (green squares),
which correspond to
the two stable isomers at the collinear configurations 
Li-CN ($\vartheta=0$\,rad, with an energy of 2281\,cm$^{-1}$) 
and CN-Li ($\vartheta=\pi$\,rad, with an energy of~0\,cm$^{-1}$).
These two minima are separated by a modest energetic barrier for the
isomerization along the minimum energy path (MEP)
given by $R_\text{MEP}(\vartheta)$ (dashed green line),
whose top has an energy of~$E_\text{SP}=3455$\,cm$^{-1}$
and a saddle$\times$center 
structure in the phase space (index-1 saddle, shown as a cross).

In the next section,
we report a simplified PES, 
which is still able to reproduce the main characteristics
of the 
\emph{ab initio} PES
introduced in Fig.~\ref{fig:1}.

\subsection{Equivalent potential energy surface with 2 degrees of freedom}
\label{sec:2dof_Vequiv}

As can be seen, for constant~$\vartheta$,
the \emph{ab initio} PES shown in Fig.~\ref{fig:1}
increases abruptly for small~$R$
while it is a 
 slow varying function for large~$R$.
This behavior is well described by 
\begin{equation} \label{eq:Vmorse2}
   V_{\rm equiv}(R, \vartheta) = V_{\rm MEP}(\vartheta) + V_{\rm Morse}(R, \vartheta),
\end{equation}
where~$V_{\rm MEP} (\vartheta) = V_{\rm{a.i.}}(R_{\rm MEP}(\vartheta),\vartheta)$ is the potential along the MEP,
and
\begin{equation} \label{eq:Vmorse}
   V_{\rm Morse}(R, \vartheta) = D(\vartheta) \left[ 1- e^{-\alpha(\vartheta) (R-R_{\rm min})} \right]^2
\end{equation}
is the Morse potential,
which depends on the distance~$R$ and
also on the angular coordinate,~$\vartheta$,
through the well depth~$D(\vartheta)$
and the width~$\alpha(\vartheta)$.
This width is usually expressed as 
a function of the
frequency~$\Omega(\vartheta)$
as~$\alpha(\vartheta) = \sqrt{\Omega(\vartheta) \mu_1/(2D(\vartheta))}$.
Consequently,
a different Morse potential is used for each~$\vartheta$.

\begin{figure}
\includegraphics[angle= 0, width=0.95 \columnwidth]{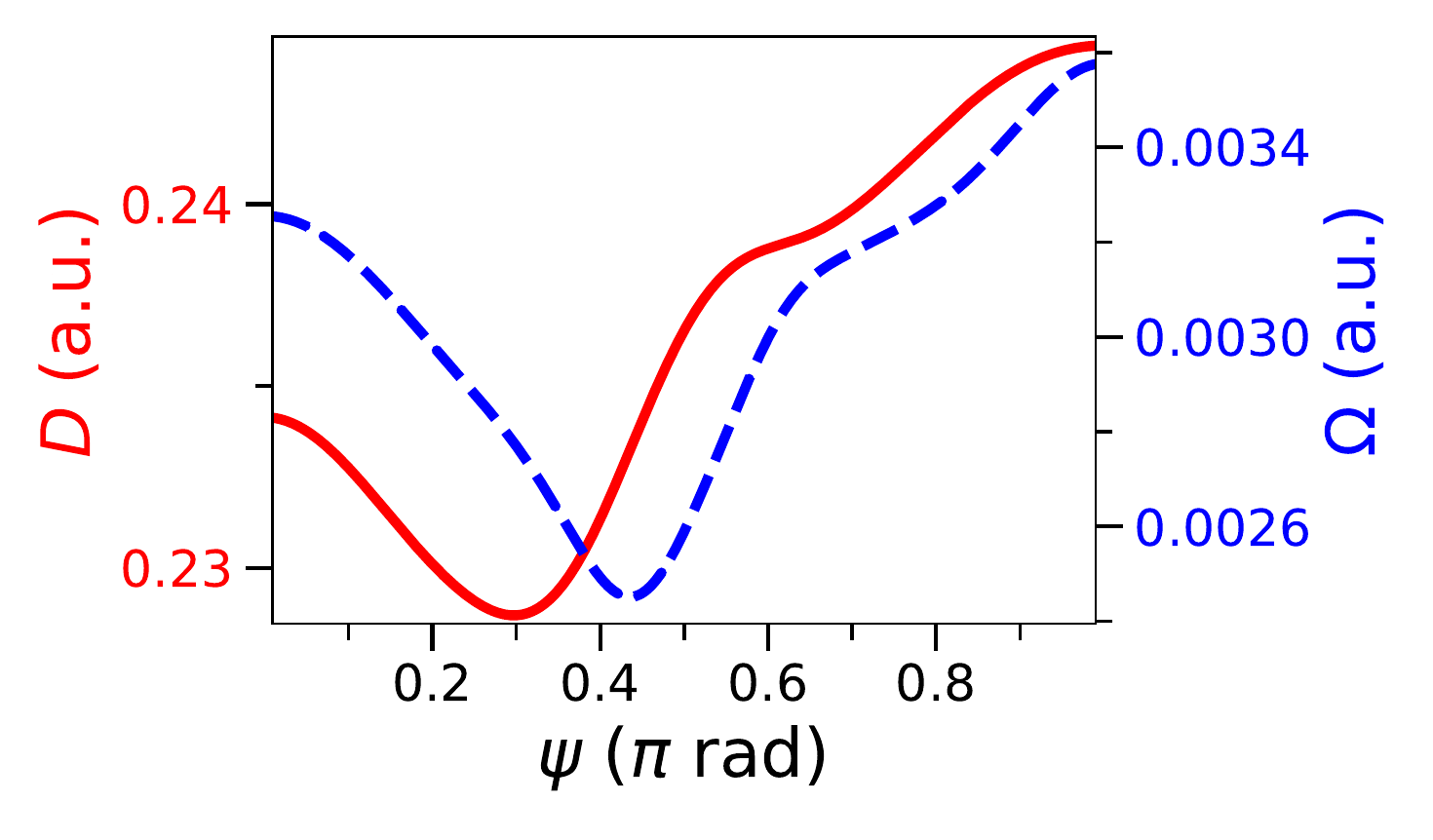}
\caption{Parameters of the Morse potential:
trap depth~\eqref{eq:D} (continuous red line),
and frequency~\eqref{eq:w} (dashed blue line).
%
}
  \label{fig:Dw}
\end{figure}

In order to be able to use the potential~\eqref{eq:Vmorse}
efficiently,
we have performed an additional fitting of the functions~$D(\vartheta)$ 
and~$\Omega(\vartheta)$ as
\begin{eqnarray} \label{eq:expan}
   D(\vartheta) &=& \sum_{n=0}^9 d_n \cos (n \vartheta), \label{eq:D} \\
   \Omega(\vartheta) &=& \sum_{n=0}^9 w_n \cos (n \vartheta). \label{eq:w}
\end{eqnarray}
The fitted coefficients~$d_n$ and~$w_n$ are listed in Table~\ref{Tab:I}.
We have verified that the error in Eqs.~\eqref{eq:D} and~\eqref{eq:w}
with respect to their original expressions
is in all cases less than 0.3\%,
which provides an estimation of the difference between the
 potential~\eqref{eq:Vmorse2} and the original
\emph{ab initio} potential.
Figure~\ref{fig:Dw} shows the 
values of~$D(\vartheta)$ 
and~$\Omega(\vartheta)$ obtained for the system under study.
As can be seen,
both functions have two maxima at~$\vartheta=0, \pi$\,rad,
and a minimum in between
(found at~$\vartheta\sim 0.30 \pi$ and~$0.44 \pi$\,rad, respectively).

In the next section we briefly summarize
the vibrational dynamics for Hamiltonian~\eqref{eq:H_2dof}.
Let us remark that the results reported there,
which are associated with the
\emph{ab initio} PES,
are also quantitatively valid for the 
adiabatically obtained PES~\eqref{eq:Vmorse2} due to their
high similarity.
 
%
\begin{table}
   \caption{Fitted coefficients $d_n (\pm \, 3 \times 10^{-7}$\,a.u.)
                  and~$w_n ( \pm \, 5 \times 10^{-8}$\,a.u.)
                   of Eqs.~\eqref{eq:D} and~\eqref{eq:w}, respectively.
}
   \label{Tab:I}
   \begin{tabular}{ccc}
    \hline
    \hline
    $n$ & $d_n$(10$^{-4}$\,a.u.)& $w_n$ (10$^{-5}$\,a.u.) \\ 
    \hline
     0 &  $2362.990 $  & $303.267 $  \\
     1 &  $-67.946 $&  $ -21.217 $ \\
     2 &  $ 17.920$ & $36.838 $  \\
     3 &  $17.416 $ & $11.288 $  \\
     4 &  $14.756 $ & $-2.689$  \\
     5 &  $-1.468$ &  $-7.388  $ \\
     6 &  $4.181  $ & $4.693$  \\
     7 & $0.845 $  &  $2.215  $ \\
     8 & $1.290 $ &  $-0.616 $ \\
     9 & $0.054$ & $-1.144  $  \\
%
    \hline
    \hline
   \end{tabular}
\end{table}

\subsection{Trajectories and vibrational dynamics}
\label{sec:dyn}

The dynamics of this system is followed by computation of classical
trajectories~$(R, \vartheta, P_R, P_\vartheta)$
which evolve in the four-dimensional phase space
as Hamiltonian~\eqref{eq:H_2dof} has 2 dof.
In practice,
all initial conditions are taken on 
an adequate
Poincar\'e surface of section (PSOS)~\cite{LL10}.
In this case, such a suitable PSOS is
choosen along the MEP,
which renders the most relevant dynamical information, 
i.e.,~that concerning the angular motion.
As the MEP is not an actual trajectory of LiCN, and in order to make
the PSOS an area preserving map, a transformation to new 
coordinates must be performed~\cite{Benito89,Borondo95,Revuelta17}
\begin{eqnarray}
  \psi & = & \vartheta, \qquad\qquad\qquad \; 
   P_\psi= P_\vartheta - \left( \frac{dR_\text{MEP}}{d\vartheta} \right)_{\vartheta=\psi} P_R \nonumber \\
  \rho & = & R-R_\text{MEP}(\vartheta),  \quad
   P_\rho = P_R.
  \label{eq:2}
\end{eqnarray}

In Fig.~\ref{fig:1}, we also plot superimposed four POs
which are relevant for this work.
In particular, the purple one (second one starting from the right)
is marginally stable, running almost vertically.
This orbit emerges 
``out of the blue''` in a tangent or saddle-node bifurcation~\cite{Borondo95,Borondo96} 
(SNB)
$SN_1$ at $E_{SN_{1}}=3440.6$~cm$^{-1}$ (just below the energy $E_\text{SP}$,
which must be exceeded in order to permit the isomerization).
Subsequently, as energy increases this PO bifurcates in a pair, 
the one that moves leftwards being stable and the one moving rightwards being unstable.
The result is shown in Fig.~\ref{fig:1} for 
$E=4000$~cm$^{-1}$ in blue and red colors,
respectively.
The corresponding manifolds and their foldings and intersections~\cite{Revuelta19} 
are very intricate, as shown in Fig.~\ref{fig:2}.
Hereafter,
we refer to the unstable PO as \SNu-PO.
Other bifurcations in this molecule~\cite{Prosmiti96,Arranz00} and other
systems~\cite{Inarrea11, Allahem12} have been similarly reported.

Let us conclude this section by remarking that SNBs are also important
quantum mechanically,
since they can give rise to the so-called \textit{superscars},
ith the result
that some wave functions are
strongly localized along the 
bifurcated POs.
This phenomenon, first studied in a quantum map~\cite{Keating01},
has also been described in the molecule under study~\cite{Borondo96, Revuelta16}.
The imprint of SNBs on spectral properties has also been 
considered by other authors~\cite{Joyeux02, Schomerus97}.

Finally, we also show in Fig.~\ref{fig:1} the leftmost (brown) PO, 
which is a non-recrossing dividing surface~\cite{Pechukas77}.
We  refer to this PO as TS-PO 
as it lies close to the position where the transition state 
or activated complex is formed~\cite{Pechukas77, GM12, Revuelta16b}.
It first appears at the saddle-point energy $E_\text{SP}$.

\section{Lagrangian descriptors} \label{sec:LD}

\begin{figure}
\includegraphics[width=0.85 \columnwidth, angle=0]{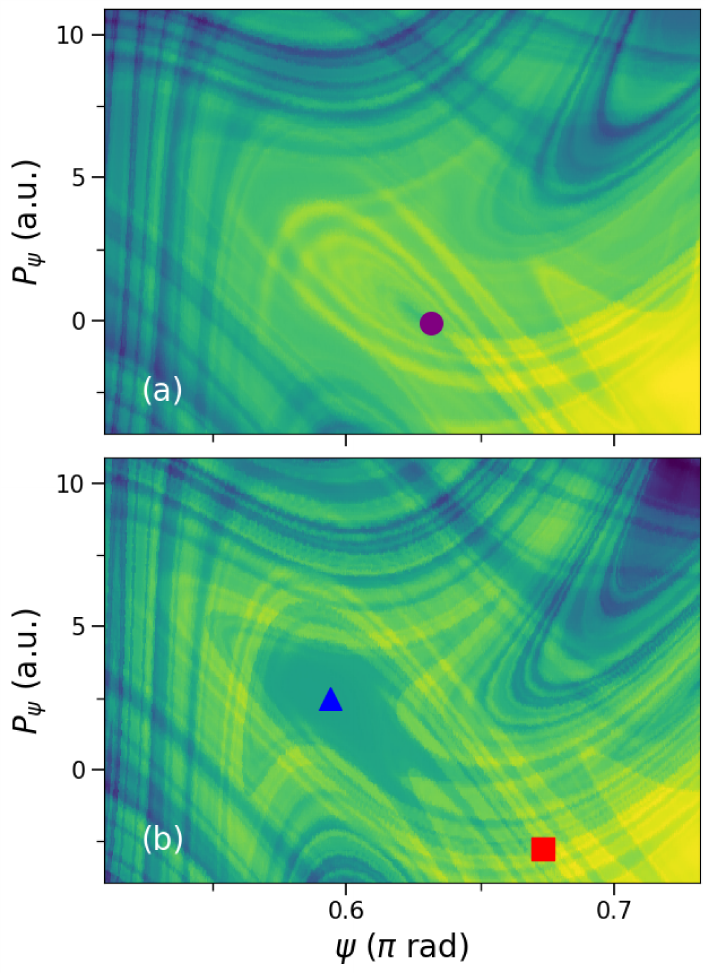}
\caption{Lagrangian descriptors as defined in Eq.~\eqref{eq:LD} 
with~$p=0.4$, and~$\tau = 2 \times 10^4$\,a.u.\ computed from trajectories of LiCN
for (a)~$E=E_{\rm SN_1} = 3440.6$\,cm$^{-1}$, and
for (b)~$E= 4000$\,cm$^{-1}$.
The purple circle, blue triangle and red square indicate the position of the 
parabolic, elliptic and hyperbolic points associated with the three rightmost
periodic orbits shown in Fig.~\ref{fig:1},
which emerge due to a saddle-node bifurcation at~$E=E_{\rm SN_1}$.}
\label{fig:2}
\end{figure}

In order to unveil the phase space structures existing in our system 
in connection with the 
POs
described in Sec.~\ref{sec:system},
we  use the LD computed as~\cite{Mancho13, Lopesino15, Revuelta19}
\begin{equation} \label{eq:LD}
    M(\mathbf{z}_0,\tau)=\int_{-\tau}^\tau \sum_{i=1}^4\vert z_i(t)\vert^p \, dt,
\end{equation}
where~$\mathbf{z}_0 = \left(\psi_0, R_0=R_\text{MEP}(\psi_0),P_{\psi,0},P_{\rho,0}=0\right)$
is a vector of initial conditions taken in the PSOS defined in Eq.~\eqref{eq:2}, 
$p$ a parameter defining the chosen norm,
and $\tau$ a parameter defining the time interval in which the LD is
calculated.
Notice that the propagation of the trajectory is done forward and backward in
order to capture the effects in the phase space of
the  unstable and stable manifolds at the same time.
Some results for Hamiltonian \eqref{eq:H_2dof} are shown in Fig.~\ref{fig:2},
both below and above the 
SNB
energy,
for~$p = 0.4$ and~$\tau=2 \times 10^4$\,a.u.~\cite{Revuelta19},
this last integration time
being 1 order of magnitude larger than the
periods of the POs of interest shown in Fig.~\ref{fig:1}.
We have also highlighted with different symbols the fixed points 
associated with those POs. 
As can be seen, the phase space is formed by a very complex structure 
due to the homoclinic intersections of the invariant manifolds,
which become visible where the value of the LDs changes
abruptly from a large value (shown in dark blue)
to a small one (in yellow).
These manifolds are identified as singularities in the LD plots, 
as is discussed later in more detail in Sec.~\ref{sec:results}.
Notice also the existence of a region where the LDs are smooth functions,
where they remain almost constant;
this region is associated with the stability island that surrounds 
the stable PO, marked as a blue triangle.
Note in this respect that LDs can also be used to characterize
invariant tori~\cite{Montes21}
and the repeller in open systems~\cite{Carlo20}.
The structure of the LDs in the whole phase space at these energies 
is shown in Fig.~A.\ref{fig:LD_LiCN_2D_p04} in Appendix~\ref{sec:2d}.

Due to the complexity of the geometrical objects in Fig.~\ref{fig:2},
we consider in the next section the results yielded by two simplified,
yet equivalent, models.
Both models are constructed using Morse potentials.
The first one is that previously reported in Sec.~\ref{sec:2dof_Vequiv};
it has 2 dof and 
reproduces the characteristics 
of Eq.~\eqref{eq:H_2dof} in the vicinity of the SNB
with great precision.
The second one has only 1 dof and was
originally introduced in 
Ref.~\onlinecite{Borondo95}.
This reduced dimensional model is still able to qualitatively 
reproduce the phase-space structures associated with  Eq.~\eqref{eq:H_2dof}
but not 
the heteroclinic
intersections that occur when the invariant manifolds associated with different
POs 
intersect.

\section{Results and discussion} \label{sec:results}
%
In this section we present our results and the corresponding discussion.
This section is divided in three parts.
First, 
we discuss in Sec.~\ref{sec:tau} the influence of the integration time
on the LDs, showing that
the invariant manifolds associated with a particular PO
require a computation time large enough compared to
the inverse of its characteristic exponents.
Second, 
as an intermediate step towards the 1-dof equivalent model,
we demonstrate the excellent performance of 
the equivalent adiabatic PES
given by Eq.~\eqref{eq:Vmorse2}.
Third, we conclude this section by 
studying the reduced dimensional model based on the
adiabatic approximation,
which is, nevertheless, capable of reproducing the main
structures that determine the dynamics of the molecule.

\subsection{Influence of the integration time}
\label{sec:tau}

Lagrangian descriptors are able to unravel
invariant manifolds 
in phase space
only if the integration time~$\tau$
appearing in Eq.~\eqref{eq:LD} is \emph{sufficiently} large
to account for their particular hyperbolic behavior. 

The exponential sensitivity 
of an initial condition found in the neighborhood of
a given unstable PO
becomes manifest on 
a characteristic time scale
given by the inverse of the
stability  exponents
of its corresponding invariant manifolds.
For systems with 2 dof, 
there are two of such exponents,~$\lambda_{s, u}$,
each one
associated with the stable and the unstable manifold,
respectively;
the 
associated eigenvalues of the monodromy
matrix are given by~$\gamma_{s,u}=e^{\lambda_{s, u} T}$,
where~$T$ is the period
and,
as~$\lambda_u = - \lambda_s  \ge 0$,
$\gamma_s \gamma_u = 1$.
In general,
a neighboring trajectory will move apart
from the reference PO at a rate given by~$\sim$\,$e^{\lambda_u t}$
in the direction of the unstable manifold.
Conversely,
it will approximate the PO
in the direction of the stable manifold
at a rate given by~$\sim$\,$e^{\lambda_s t}$
(or separate from it at the same rate 
when evolving backwards in time).
Consequently,
the particular character of the manifolds is
expected to show up only for~$\vert \tau \vert \ge \vert  \lambda_{u, s}^{-1}\vert $.

\begin{figure}
\includegraphics[angle= 0, width=0.85 \columnwidth]{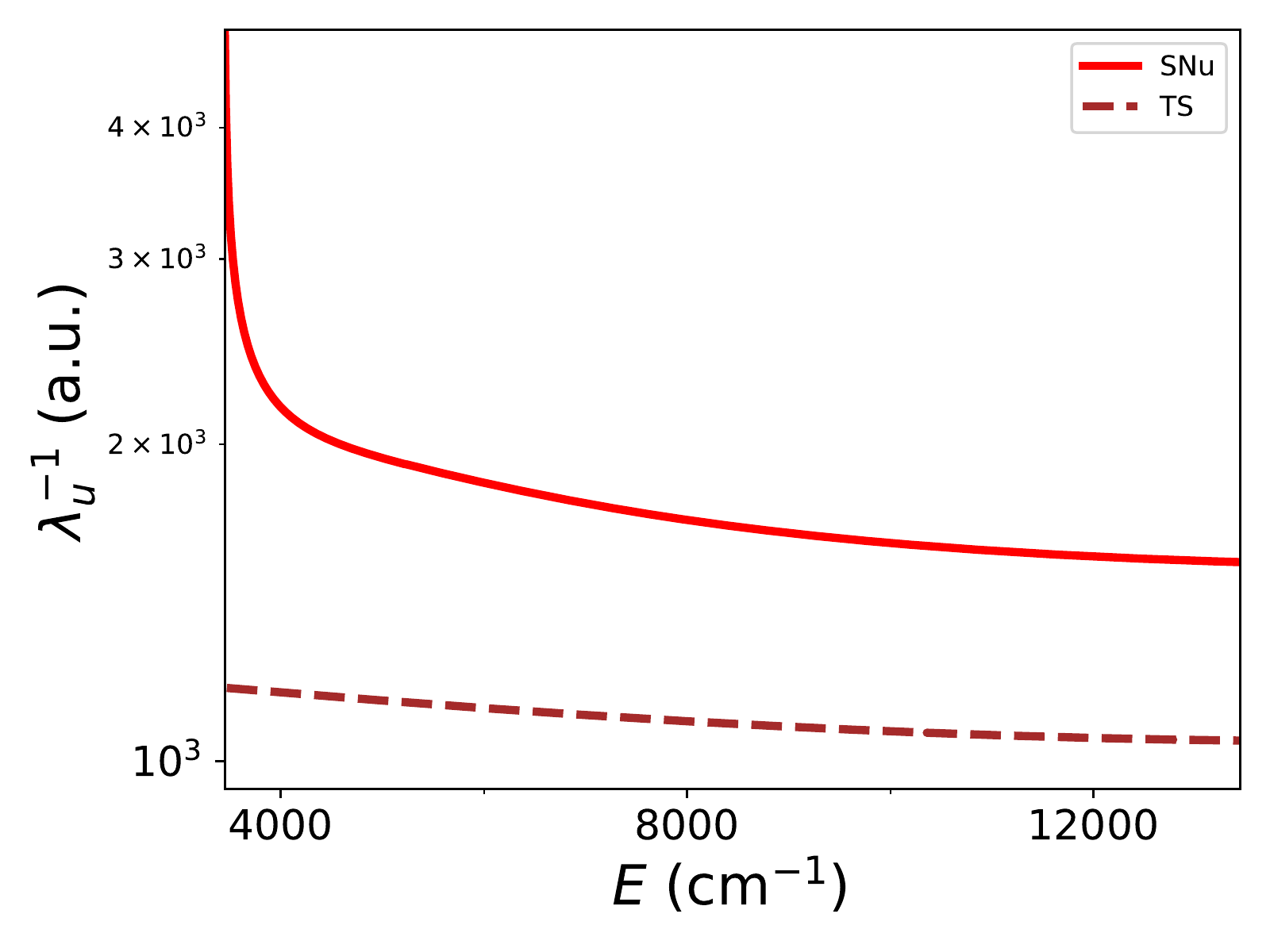}
\caption{Inverse of the unstable stability exponent as a function of the energy.
The top red (bottom brown) curve corresponds to
the unstable periodic orbit
(transition state trajectory)
 shown with the same coloring in Fig.~\ref{fig:1}.}
%
%
  \label{fig:invlambda}
\end{figure}

In  this work,
we are mostly interested in the 
invariant manifolds associated with the 
unstable  (left brown) TS-PO and
(right red) \SNu-PO
presented in Fig.~\ref{fig:1},
which are responsible for the formation barriers
that obstruct isomerization.
These two POs have different stability exponents,
as can be seen in Fig.~\ref{fig:invlambda},
where the inverse of their unstable exponent
is shown.
Notice that in both cases,
$\lambda_u^{-1}$ reduces with the energy,
though  in the case of the \SNu-PO
this happens in a much more dramatic way,
especially at small energies.
Actually,~$\lambda_{{\rm SN}u-{\rm PO}, u}^{-1}$ diverges 
at
$E=E_{\rm SN_1} = 3440.6$\,cm$^{-1}$,
since at the bifurcation this unstable PO collapses
with the (blue) stable PO,
rendering the (purple) marginally stable PO,
whose monodromy matrix has two eigenvalues equal to 1,
so its characteristic exponents must 
cancel (and then~$\lambda_{u, s}^{-1} \rightarrow  \pm \infty$).
%

\begin{figure}
\includegraphics[angle= 0, width=0.85 \columnwidth]{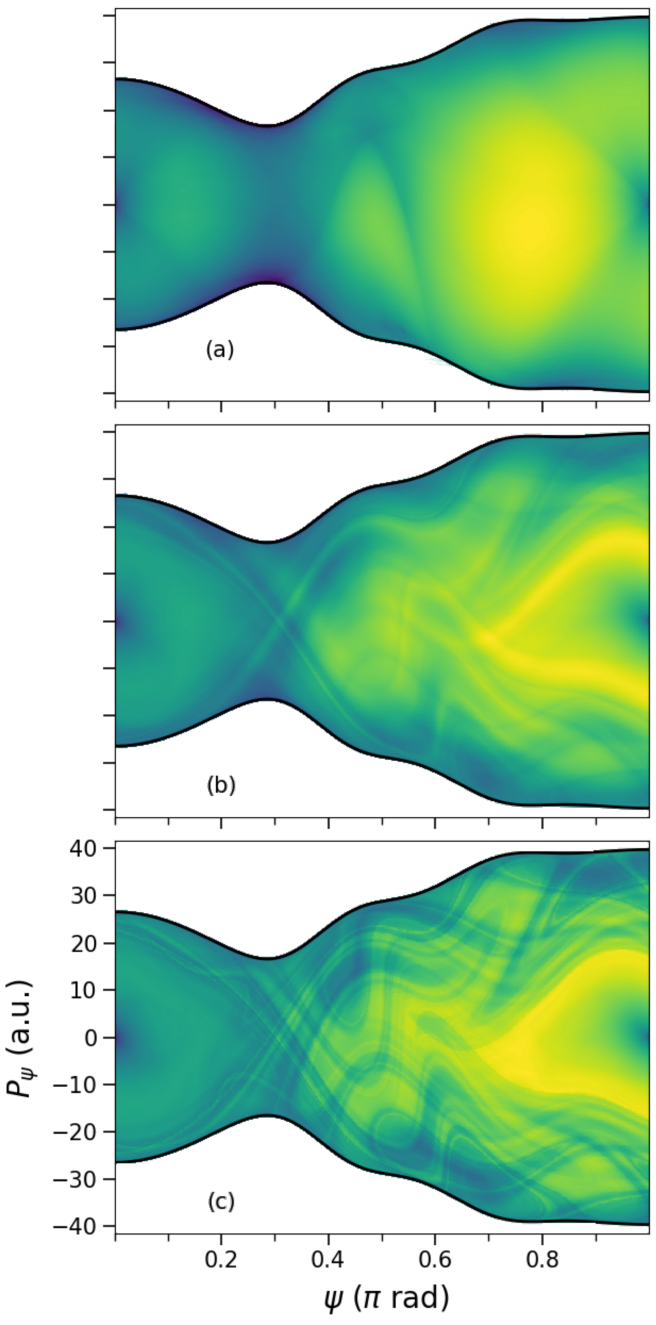}
\caption{Lagrangian descriptors as defined in Eq.~\eqref{eq:LD} 
computed from trajectories of LiCN
for~$E= 4000$\,cm$^{-1}$,~$p=0.4$, 
and~$\tau = C \lambda_u^{-1}$, 
$\lambda_{\textnormal{TS-PO},u}^{-1} \simeq 1163$\,a.u. 
being the inverse of the stability exponent of the TS trajectory
shown in brown in Fig.~\ref{fig:1},
and~$C=1$ (a), 5 (b), and~10 (c), respectively.}
\label{fig:LDinvlambda_TS}
\end{figure}

\begin{figure}
\includegraphics[angle= 0, width=0.85 \columnwidth]{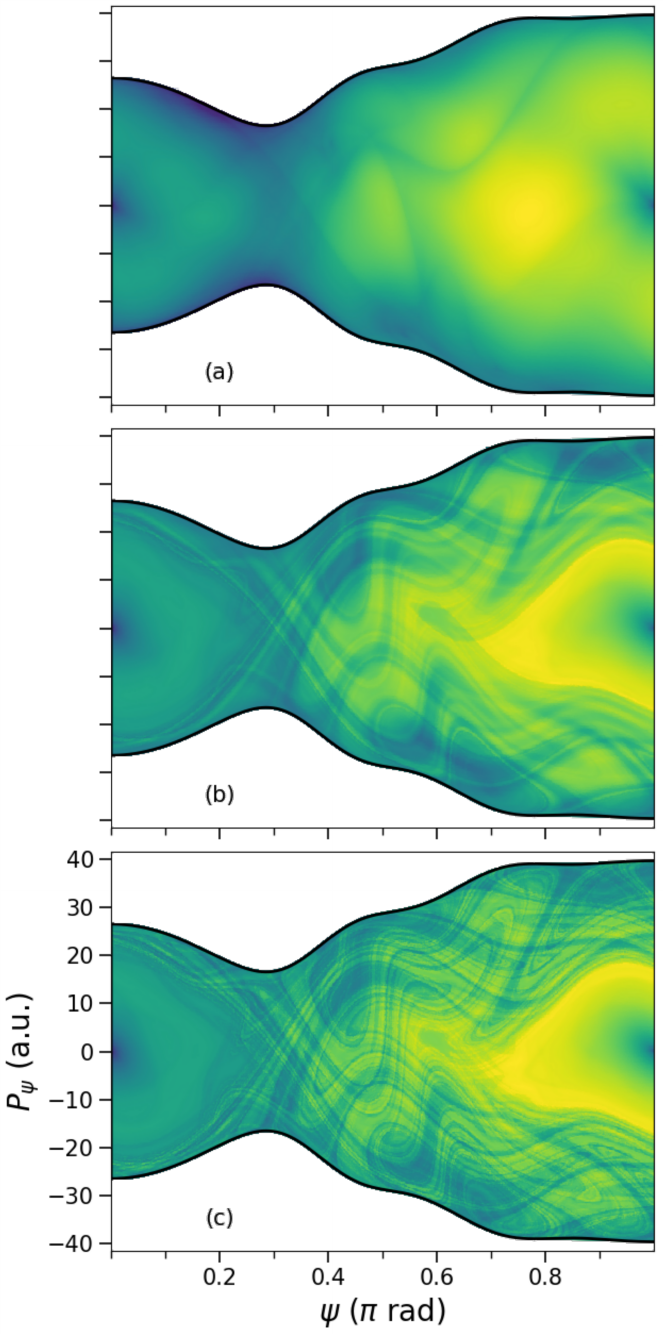}
\caption{Same as Fig.~\ref{fig:LDinvlambda_TS}
for the {\SNu} trajectory
shown in red in Fig.~\ref{fig:1},
whose stability exponent has an inverse
equal to~$\lambda_{\textnormal{SN}_\textnormal{u}-\textnormal{PO}, u}^{-1} \simeq 2170$\,a.u.
at the energy of~$E=4000$\,cm$^{-1}$.} 
\label{fig:LDinvlambda_SNu}
\end{figure}

Figures~\ref{fig:LDinvlambda_TS}
and~\ref{fig:LDinvlambda_SNu} show the value of
the LDs as defined in Eq.~\eqref{eq:LD} for
different characteristic times~$\tau = C \lambda_u^{-1}$
as a function of the stability exponent of the TS-PO
and the \SNu-PO, respectively.
As in Fig.~\ref{fig:2},
dark blue (yellow) coloring indicates a large (small)
value in the LDs.
Note that for the used energy of~$E=4000$\,cm$^{-1}$
the stability exponent for the TS-PO is almost
two times larger than that of the \SNu-PO.
Consequently, 
the integration times chosen in 
Fig.~\ref{fig:LDinvlambda_TS}
are approximately half of those used in
Fig.~\ref{fig:LDinvlambda_SNu}.
This is the reason why 
Fig.~\ref{fig:LDinvlambda_SNu} 
has a more detailed structure
than Fig.~\ref{fig:LDinvlambda_TS}.

When the integration time is small
[by taking, for example, $\tau=\lambda_u^{-1}$,
as done in Figs.~\ref{fig:LDinvlambda_TS}(a)
and~\ref{fig:LDinvlambda_SNu}(a)],
the LD-plots show up as a blurry picture. 
Nonetheless,
the results for Fig.~\ref{fig:LDinvlambda_SNu}(a)
start to unveil the
structure around the TS-PO,
contrary to what happens in
Fig.~\ref{fig:LDinvlambda_TS}(a),
where the integration time is still too small.
Then,
larger integration times are required to allow the
identification of these invariant manifolds.
Notice in particular the situation
shown in Fig.~\ref{fig:LDinvlambda_TS}(b),
where~$\tau=5 \, \lambda_{{\rm TS-PO}, u}^{-1}$.
There,
the manifolds emanating from the TS-PO
are clearly visible,
but not those associated with the \SNu-PO,
because the characteristic exponent for this PO
is smaller than for the previous one, 
and then longer times are required to study its behavior.
The stability island that lies close to this trajectory in
nevertheless visible.
The manifolds for the \SNu-PO
can be seen for a larger integration time
such as~$\tau=5 \, \lambda_{{\rm SN_\textnormal{u}-PO}, u}^{-1}$
as inferred by inspection of Fig.~\ref{fig:LDinvlambda_SNu}(b).

To conclude,
we show in Figs.~\ref{fig:LDinvlambda_TS}(c)
and~\ref{fig:LDinvlambda_SNu}(c)
the results for an integration time that is $\tau = 10 \, \lambda_u^{-1}$.
In both cases, but especially in the second one,
a very detailed picture of the chaotic region of phase space is obtained,
where the complex structure of the heteroclinic tangle
becomes visible.
Thus,
in the rest of the article
integration times equal to
$\tau = 2 \times 10^{4}$\,a.u. will be used,
a value that is slightly smaller than that considered 
in Fig.~\ref{fig:LDinvlambda_SNu}(c).
Similar results are obtained for other comparable
integration times 
as long as they are large \emph{enough}
compared to~$\lambda_u^{-1}$.
Let us conclude by pointing out that this 
criterion is
not 
applicable to
marginally stable POs
(as it happens for the \SNu-PO at~$E=E_{\rm SN_1}$),
since then~$\lambda_{u, s}=0$, 
and, as a consequence,~$\lambda_{u, s} \rightarrow \pm \infty$.

\subsection{Equivalent model with 2 degrees of freedom}
\label{sec:Vequiv2dofs}
In this section we discuss the excellent performance of 
the alternative PES  given by Eq.~\eqref{eq:Vmorse2}.
As the original \emph{ab initio} PES,
the new one has 2 dof
so the system dynamics takes place in a 4-dimensional phase space.
This study is conducted as a necessary intermediate
step in order to to develop the
reduced dimensional model reported in 
Sec.~\ref{sec:1dof}.
%

\begin{figure}
\includegraphics[width=0.85 \columnwidth]{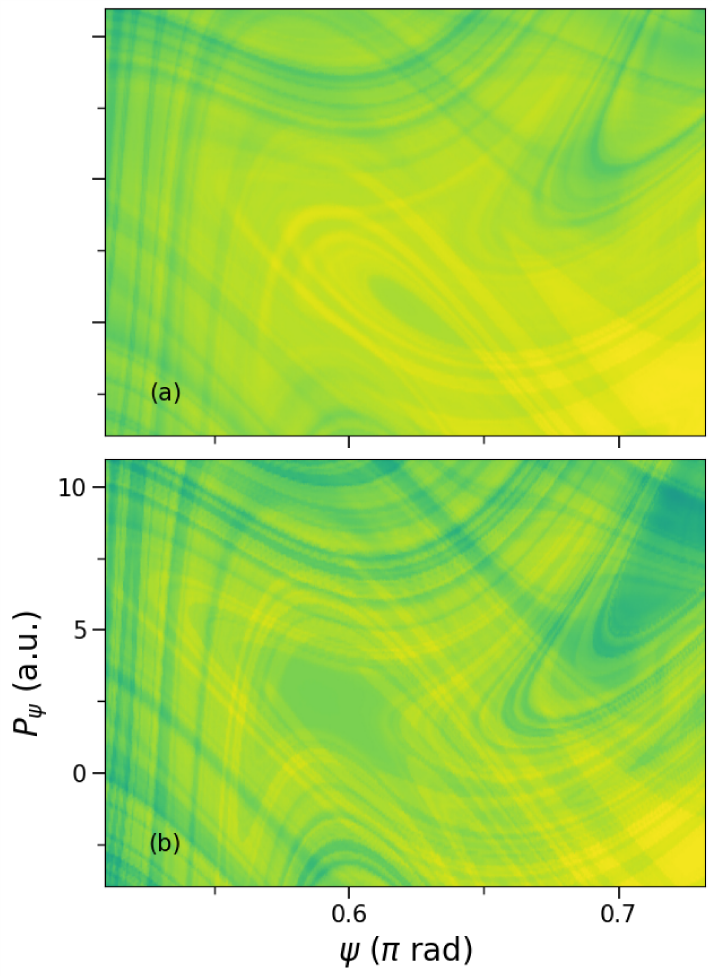}
\caption{Same as Fig.~\ref{fig:2} for the
equivalent two-dimensional potential energy surface
constructed using the adiabatic Morse potential.
In both case, the exact moment of inertia~$\mathcal{I}_\vartheta 
= [ 1/(\mu_1 R^2) + 1/(\mu_2 r_{\rm eq}^2)]^{-1}$
is used.
}
  \label{fig:LD_Morse_2D_Ivar}
\end{figure}
\begin{figure}
\includegraphics[width=0.85 \columnwidth]{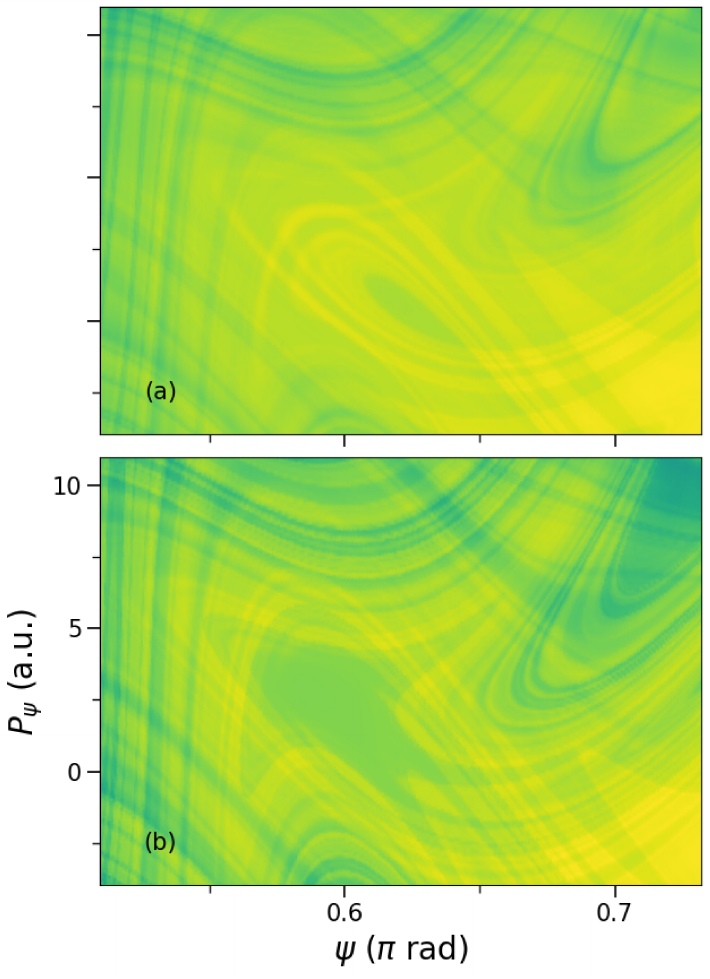}
\caption{Same as Fig.~\ref{fig:LD_Morse_2D_Ivar}
for a constant
moment of inertia~$\mathcal{I}_{\rm SP} =
[ 1/(\mu_1 R_{\rm MEP}^2(\vartheta_{\rm SP})) + 1/(\mu_2 r_{\rm eq}^2)]^{-1} 
\simeq 3.94 \times 10^4 $\,amu.
}
  \label{fig:LD_Morse_2D_cons}
\end{figure}

Figure~\ref{fig:LD_Morse_2D_Ivar} shows the value of the LDs
in the vicinity of the SN bifurcation
for the same set of parameters as those previously used in Fig.~\ref{fig:2}
but modelling the PES with Eq.~\eqref{eq:Vmorse2}.
As can be seen,
the structure that is observed 
using any of the two PESs
is similar,
both below [(a) panels]
and above [(b) panels] the bifurcation energy~$E_{\rm SN_1}$.
Notice that the results shown in 
Figs.~\ref{fig:2} and~\ref{fig:LD_Morse_2D_Ivar}
have been obtained with the Hamiltonian~\eqref{eq:H_2dof},
which has a moment of 
inertia
associated with the angular coordinate,~$\mathcal{I}_\vartheta = [ 1/(\mu_1 R^2) + 1/(\mu_2 r_{\rm eq}^2)]^{-1}$,
that is $R$-dependent.
As a further simplification,
one can take this moment of inertia as constant
by setting it, e.\,g., equal to its value at the top of the 
largest energetic barrier,~$\mathcal{I}_{\rm SP} = [ 1/(\mu_1 R_{\rm MEP}^2(\vartheta_{\rm SP})) + 1/(\mu_2 r_{\rm eq}^2)]^{-1} \simeq 4 \times 10^4$\,amu.
As shown in Fig.~\ref{fig:LD_Morse_2D_cons}(a),
the structure of the phase space is almost equal
to those already presented in 
Figs.~\ref{fig:2}(a) and~\ref{fig:LD_Morse_2D_Ivar}(a),
while above the bifurcation  energy
only minor differences are 
visible, 
as inferred by comparison of
Figs.~\ref{fig:2}(b),~\ref{fig:LD_Morse_2D_Ivar}(b),
and~\ref{fig:LD_Morse_2D_cons}(b).

The agreement 
between the results for the \emph{ab initio} PES
and those associated with the Morse-based PES
allows us to make a further simplification
in order to define a new model with only 1 dof,
as discussed in the next section.
%
%

\subsection{Equivalent model with 1 degree of freedom}
\label{sec:1dof}

All POs shown in Fig.~\ref{fig:1} correspond to almost 
pure
vibrational stretching states,
where the distance between 
the Li atom and the CN fragment changes 
periodically, while the angle~$\vartheta$ remains almost constant.
In this situation an adiabatic separation between the two modes
can be carried out with good approximation, and then we neglect the
stretching motion.
Accordingly, Hamiltonian~\eqref{eq:H_2dof} can be substituted 
by the following 
1-dof expression
\begin{equation}  \label{eq:H_1dof}
  \mathcal{H}_1 = \frac{P_\psi^2}{2\mathcal{I}_\psi} 
                  + V_\text{eff} (\psi),
\end{equation}
where~$(P_\psi, \psi)$ are now the 
(two-dimensional)
phase-space coordinates;
$ \mathcal{I}_\psi$ is the moment of inertia associated with the coordinate~$\psi$,
which will be taken equal to its value at the saddle point located at the 
energy barrier top, i.e.,~$I_\psi = I_{\rm SP}$;
and~$V_\text{eff}$ is an effective potential for~$\psi$.
The accuracy of this model has been also
assessed in
previous works~\cite{GM12, Revuelta16b, Junginger16}.
In this section,
we  demonstrate 
in detail its ability to 
adequately reproduce the invariant manifolds
associated with the TS-PO and \SNu-POs.
The section is divided in three parts.
First, we introduce the potential energy considered
in this case, 
which depends solely on the angle~$\psi$.
Second,
Sec.~\ref{sec:e<ebif} is devoted to the analysis of the 
existing manifolds below the bifurcation energy~$E_{\rm SN_1}$,
where the \SNu-PO appears.
Third,
we conclude in Sec.~\ref{sec:e>ebif}
by discussing the situation where
the energy is larger than~$E_{\rm SN_1}$.

\subsubsection{Potential energy}
\label{sec:1dof_Vmep}

\begin{figure}
\includegraphics[width=0.85 \columnwidth]{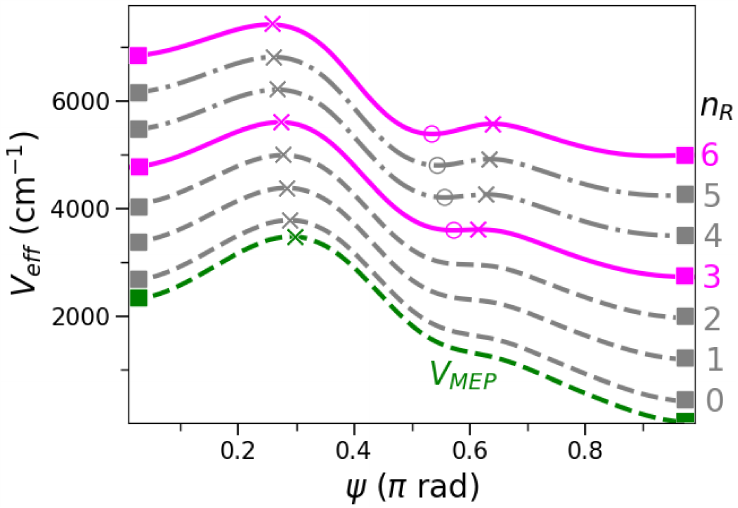}
\caption{Effective potential 
with one degree of freedom along the angular coordinate for the LiCN molecule.
The bottom dashed green line corresponds to the energy profile
along the minimum energy path, $V_\text{MEP}(\psi)$, 
of the potential energy surface shown in Fig.~\ref{fig:1}.
The remaining curves are the adiabatic potentials given by Eq.~\eqref{eq:Veff}
for the $n_R$ values shown on the right.
The positions of the minima potential wells', 
local minima,
and 
local maxima
have been 
highlighted with squares, circles, and crosses, respectively.
}
  \label{fig:Veff}
\end{figure}

The simplest approximation for~$V_\text{eff}$ in Eq.~\eqref{eq:H_1dof}
is simply given by the potential energy along the MEP,
$V_\text{MEP}(\psi)$,
shown as a dashed green line in Fig.~\ref{fig:1}.
This function is shown in the bottom dashed green line of Fig.~\ref{fig:Veff}.
As can be seen, it reproduces qualitatively the same characteristics of the 2-dof PES
$V_{\rm{a.i.}}(R,\vartheta)$, namely, two minima at $\psi = 0$ and~$\pi$~rad separated by 
a single maximum located at the barrier top $\psi= 0.29\pi$~rad.

The model described by Eq.~\eqref{eq:H_1dof} is not able to account for the 
emergence of the stable region shown in Fig.~\ref{fig:2}(b)
for~$V_\text{eff} (\psi) = V_{\rm MEP}(\psi)$
(see discussion in Sec.~\ref{sec:e<ebif}).
As discussed in Ref.~\onlinecite{Borondo95}, 
this stabilization can be explained  using
an effective potential
that is valid when the motion in the~$R$ radius
is much faster than that in the~$\vartheta$, i.\,e.,~$\psi$, angle,
just like in the stretching POs visible in the rightmost part of Fig.~\ref{fig:1}.
Then,
one can quantize the potential~$V_{\rm equiv}(R, \vartheta)$~\eqref{eq:Vmorse2}
for each value of the angle,
to define a new 1-dof potential energy as
\begin{eqnarray}   \label{eq:Veff}
   V_{\rm eff}(\psi) &=& V_{\rm MEP}( \psi ) 
                          + \hbar\Omega(\psi)\left(n_R+\frac12\right)
                          \nonumber \\
                     &-& \frac{\hbar^2\Omega^2(\psi)}{4D(\psi)}
                         \left(n_R+\frac12\right)^2,
\end{eqnarray}
where~$n_R$ is the corresponding vibrational excitation number,
and~$D=D(\psi)$ and~$\Omega=\Omega(\psi)$
are the Morse parameters given by Eqs.~\eqref{eq:D}
and~\eqref{eq:w},
which depend on the angle~$\psi$.

The effective potential~\eqref{eq:Veff} is shown in Fig.~\ref{fig:Veff}
for~$n_R=0-6$.
The potentials for~$n_R=0, 1$, and~2 (dashed gray lines)
are very similar to that for the MEP.
Consequently, the phase space presents qualitatively the same structure,
as can be seen in Figs.~B.\ref{fig:nR0}-B.\ref{fig:nR2} in Appendix~\ref{sec:nR}.
Notice, however, that the potential is flatter for~$n_R=2$ than for~$n_R=0$
around~$\psi \simeq 0.6 \pi$\,rad.
Furthermore, for~$n_R > 2$ the potential starts to show a minimum around that point,
which is precisely responsible of the stabilization process 
previously described in the discussion of Fig.~\ref{fig:1}.

%
\begin{figure}[hbtp!]
\includegraphics[width=0.85 \columnwidth, angle=0]{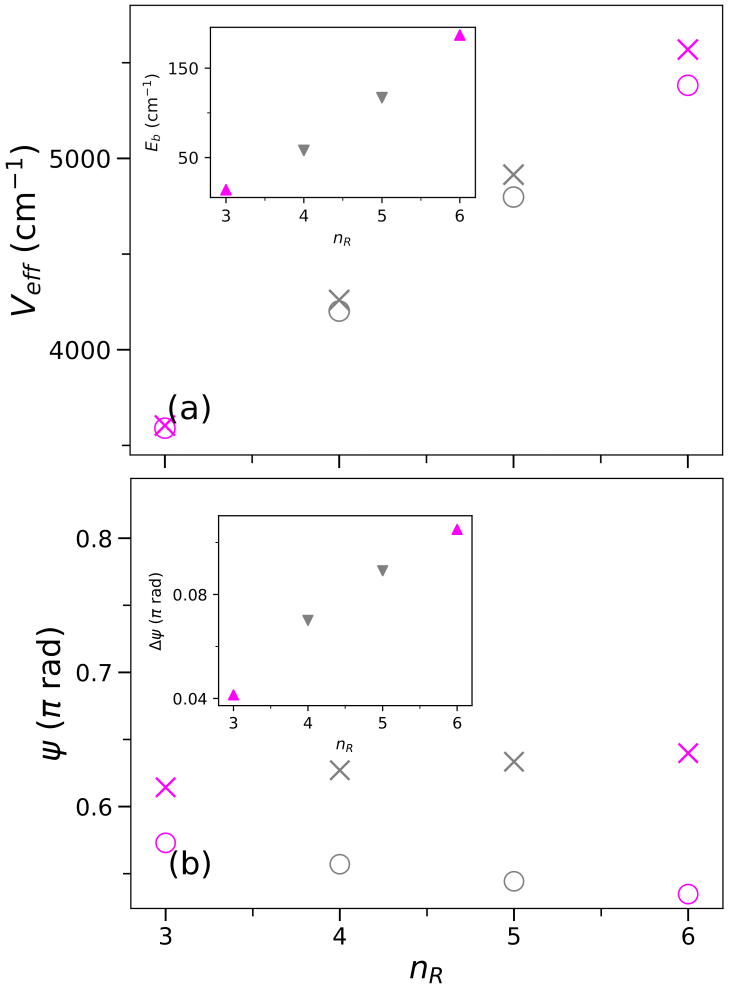}
\caption{
Characteristic parameters of the 
effective potential~\eqref{eq:Veff}
shown in Fig.~\ref{fig:Veff}.
(a) The top crosses
and empty circles show,
respectively, the potential energy at
the local maximum (saddle point in phase space) and at the local minimum (center)
of the
dynamical barrier in the LiCN molecule
as a function of the vibrational number~$n_R$.
The inset shows the height of the energetic barrier that is formed
as the difference of the results of the main panel.
(b) Position of the  local maximum
(crosses)
and of the local minimum
(empty circles) for the dynamical barrier. 
The inset shows the distance between the previous points.
}
\label{fig:barriers}
\end{figure}

Figure~\ref{fig:barriers} shows some characteristic parameters
of the effective potential~\eqref{eq:Veff}. 
Firstly,
Fig.~\ref{fig:barriers}(a) shows the value of the potential at the 
local maximum
(saddle  point) and at the 
 local minimum
(center) for the dynamical barrier. 
As can be seen, the potential energy increases with~$n_R$ for all the previous points.
However, the value of the potential at the saddle point increases faster 
than at the center, and, as a consequence, the energetic barrier height increases, 
as shown in the inset.
For example,
the energetic barrier is only~14\,cm$^{-1}$ in height for~$n_R=3$,
while it equals~187\,cm$^{-1}$ for~$n_R=6$.
Secondly,
we show in Fig.~\ref{fig:barriers}(b) the positions of the critical points
of the previous barrier.
As can be seen in the corresponding inset,
the distance between the 
saddle points and the centers 
increases with~$n_R$,
and then so does the width of the stability regions that 
show up. 
 

To conclude, let us indicate that
an alternative adiabatic approximation was obtained by  Light  and
Bačić in Ref.~\cite{Light87}, but was meant for a different purpose.
There, the authors were interested in the development of
an optimal basis set for the computation of the system eigenfunctions.
No differences should be expected at the quantum level
(eigenenergies, eigenfunctions, Husimi distributions, \ldots)
when comparing their results with those derived from
our adiabatic PES with 2 dof. 
Nevetherless,
their reduced dimensional potential energy functions
with 1 dof 
do not present any relative minimum
(see Fig.~3 of Ref.~\cite{Light87})
able to reproduce the SN bifurcation that is observed for the
system with 2 dof,
which has a strong classical imprint and a quantum imprint on 
the system \cite{Borondo95, Borondo96, Zembekov97}.

\subsubsection{Phase-space geometry below the bifurcation energy~$E_{SN_1}$} 
   \label{sec:e<ebif}

\begin{figure}
\includegraphics[width=0.85 \columnwidth, angle=0]{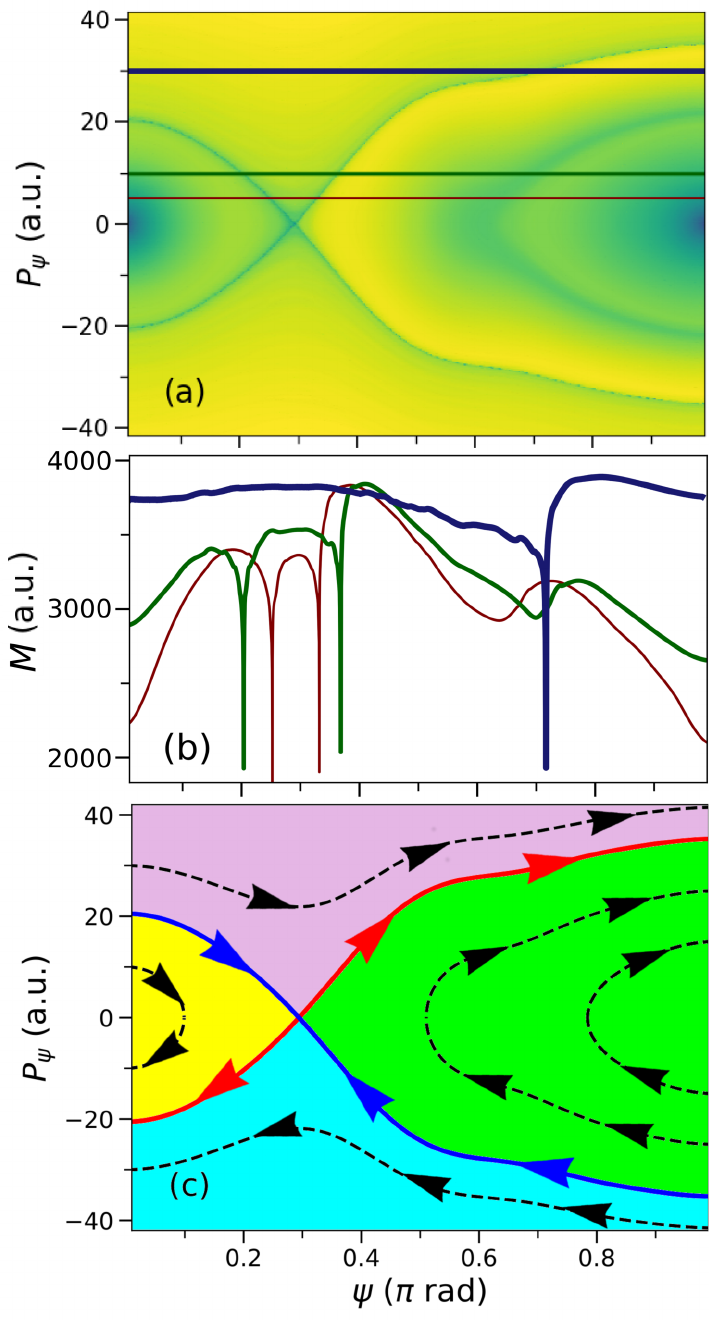}
\caption{Phase space for the 1-degree-of-freedom model of LiCN~\eqref{eq:H_1dof}
  with $V_\text{eff}(\psi)=V_\text{MEP}(\vartheta)$.
  (a) Lagrangian descriptor as defined in Eq.~\eqref{eq:LD}.
  (b) The sections of constant~$P_\psi=5, 10,$ and~30\,a.u. of the 
      Lagrangian descriptors shown in panel (a) showing singularities at the
      positions of invariant manifolds.
  (c) The stable (blue) and unstable (red) invariant manifolds associated with
      the saddle fixed point 
      of the potential
      [$(\psi, P_\psi)=(0.29\pi\,\textnormal{rad},0\,\textnormal{a.u.})$].
      They partition the phase space in four characteristic regions
     (yellow, green, purple, and cyan), 
     where the orbits (in dashed black lines) have 
     librational (embedded in the yellow and green regions) or rotational
     (contained in the purple and cyan areas) motion.
}
\label{fig:mep}
\end{figure}
Figure~\ref{fig:mep}(a) shows the LDs obtained with Eq.~\eqref{eq:H_1dof}
on the phase space for the model.
As can be seen, the LDs define a smooth function in most of the phase space.
However, there is an~``$\times$'' structure at the maximum of the $V_\text{eff}$
potential function ($\psi = 0.29 \pi$\,rad).
This structure is formed by the
invariant manifolds or separatrices emanating 
from the saddle fixed 
point found at the top of the barrier,
and it becomes visible because of the singularities that the LDs
present along the manifolds,
which are responsible for abrupt changes in the LD-plots.
This fact is more clearly illustrated in Fig.~\ref{fig:mep}(b),
where the LDs along the three horizontal lines, i.e.,~constant $P_\psi$, 
indicated in Fig.~\ref{fig:mep}(a) are plotted.
There,
conspicuous singularities are clearly observed 
when the separatrices emanating from the saddle 
point are crossed.
Indeed, for~$P_\psi = 30$\,a.u.\ (horizontal line in blue) 
there is only one such singularity,
while the brown and green horizontal lines,
corresponding to~$P_\psi = 5$ and~10\,a.u., respectively,
show two of these singularities.
This result is very interesting since 
it allows one to numerically
reconstruct using LDs
the separatrices, 
and locate the position of the parent fixed point (at their crossing), 
as it has been done in Fig.~\ref{fig:mep}(c).
Similarly to what happens in the standard pendulum~\cite{LL10},
these invariant curves separate the regions of librations and rotations, 
which in our model correspond to vibrations of the Li-CN isomer 
(left yellow region), vibrations of the Li-NC isomer (right green region),
isomerizing Li-CN$
\rightarrow 
$Li-NC trajectories (top purple region),
and isomerizing Li-NC$\leftarrow
$Li-CN trajectories (bottom cyan region).
Some examples of trajectories associated with the previous motions
have been also 
included in black dashed lines 
in Fig.~\ref{fig:mep}(c).
Let us finally remark the 
interesting 
results shown in the range $\psi\in[0.6\pi,0.7\pi]$~rad
 in Fig.~\ref{fig:mep}(b), where the green and red 
LD curves show the least abrupt minima.
This region corresponds to that where the dynamical barrier is 
formed due to the approximate inflection point in the MEP,
as discussed at the end of Sec.~\ref{sec:e>ebif}.

\subsubsection{Phase-space geometry above the bifurcation energy $E_{SN_1}$} 
  \label{sec:e>ebif}

\begin{figure}[hbtp!]
\includegraphics[width=0.85 \columnwidth, angle=0]{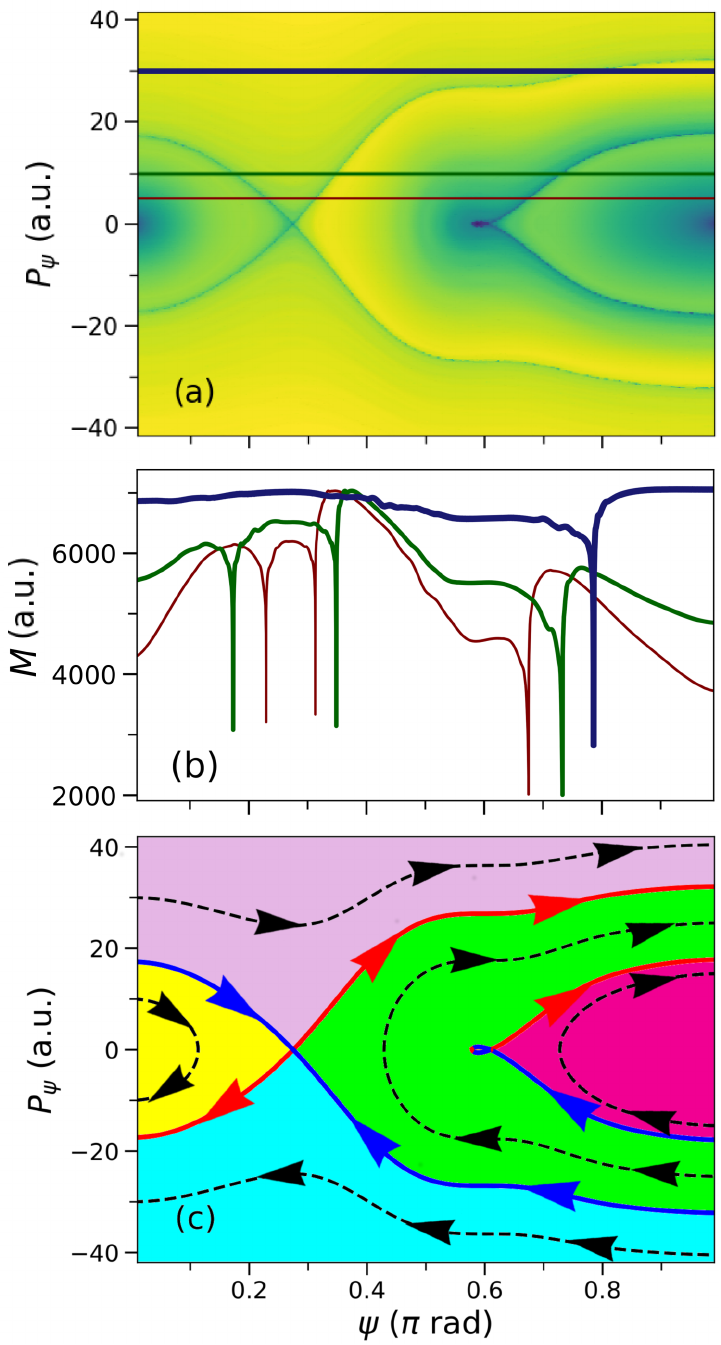}
\caption{Same as Fig.~\ref{fig:mep}
for the adiabatic potential~\eqref{eq:Veff} with~$n_R=3$.
Here, the existence of additional manifolds, associated with 
the 
saddle 
point of the secondary barrier that appears
out of the blue in a saddle-node bifurcation,
further partitions the phase space in two additional regions.
}
\label{fig:nR3}
\end{figure}

Figure~\ref{fig:nR3}(a) shows the value of the LDs
for the effective potential~\eqref{eq:Veff} with~$n_R=3$,
which is the first value of~$n_R$ where a minimum is observed.
As can be seen, the LDs are in this case also able to identify 
the invariant manifolds associated with the 
saddle point,
which do not change significantly with respect to those
shown in Fig.~\ref{fig:mep}(a), 
which are associated with an
effective potential equal to that along the MEP,
as can be inferred by visual inspection of the red
(unstable invariant manifold) and blue (stable invariant manifold)
lines in Fig.~\ref{fig:nR3}(c)
[cf.~Fig.~\ref{fig:mep}(c)].
Also, here these separatrices show up in the LD-plot as singularities,
as it becomes clearly visible
in Fig.~\ref{fig:nR3}(b), 
where the values of the LDs along the three sectioning horizontal 
lines marked in Fig.~\ref{fig:nR3}(a) are presented.
Notice however, that in this case there are additional singularities, 
which show the existence of new invariant manifolds.
These manifolds, also shown in Fig.~\ref{fig:nR3}(c),
are associated with the 
 saddle 
point 
that appears in the secondary barrier localized 
at~$\psi \simeq 0.61 \pi$\,rad. 
Notice that these manifolds, contrary to those associated with 
the 
saddle 
point discussed in the previous section,
have a different structure:
the left branch of the unstable manifold 
coincides with the left branch of the stable one:
they represent a homoclinic orbit.
This orbit encloses a stability region,
which the trajectories inside it cannot escape,
rendering a vibrational 
motion around the local minimum ($\psi \simeq 0.57\pi$\,rad).
In this case, however, the stability region is so tiny that 
the Li atom 
would describe such a small motion that it can be 
regarded as if it were fixed.
%
The right branches of the stable and the unstable manifolds emanating 
from this second saddle point also partition the phase space,
but in this case only the region that is associated with librations 
around the CN-Li isomer.
We have also shown in Fig.~\ref{fig:nR3}(c) five characteristic trajectories.
Similar comments about Fig.~\ref{fig:mep}(c) apply here.
%

\begin{figure}[hbtp!]
\includegraphics[width=0.85 \columnwidth, angle=0]{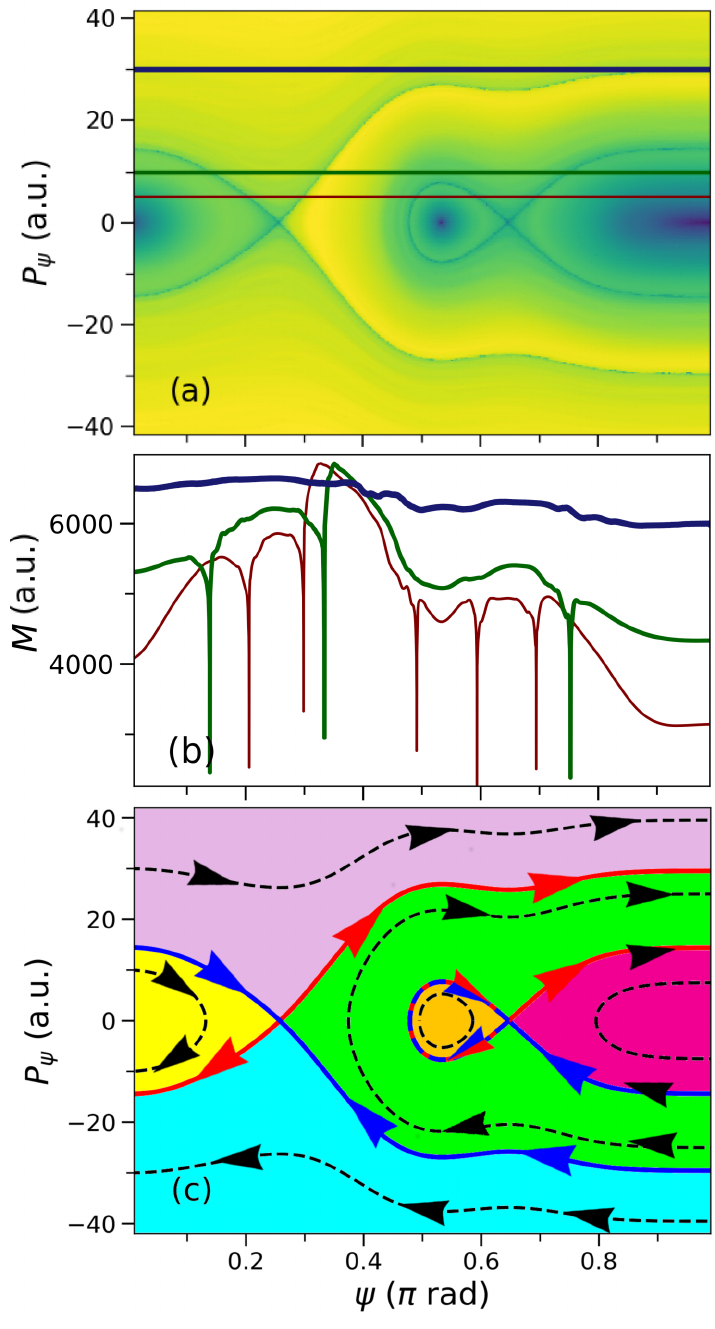}
\caption{Same as Fig.~\ref{fig:nR3} but for~$n_R = 6$.
}
\label{fig:nR6}
\end{figure}

The width of the new stability region 
shown 
 in Fig.~\ref{fig:nR3} 
increases with the integer~$n_R$.
This fact can be seen in Fig.~\ref{fig:nR6},
where a similar plot for~$n_R = 6$ is shown.
The phase-space structure is essentially the same,
but with a larger stable region.
Apart from the POs similar to those already
presented in Figs.~\ref{fig:mep}(c) and~\ref{fig:nR3}(c),
we also show in Fig.~\ref{fig:nR6}(c) a PO in the stable region,
which has the topological shape of a circle.
It represents a very particular situation
where 
the Li atom describes a rotation around the
C-N at the local minimum ($0.53 \pi$\,rad) that appears through a
dynamical process.
The evolution of the phase space for $n_R = 4, 5, 7, 8$ and~9
can be found in Figs.~B.\ref{fig:nR4}-B.\ref{fig:nR9} 
of the Appendix~\ref{sec:nR}, respectively.
As already mentioned, 
the area of the stable region increases with the integer~$n_R$.
This fact agrees with the previous discussion on Fig.~\ref{fig:barriers},
where it was shown that the height of the local barrier and
the distance between the local minimum and the maximum
that determines the 
basin size increase with~$n_R$.
%

\section{Summary and outlook} \label{sec:summary}
In this paper, we have applied the Lagrangian descriptors to identify
the invariant manifolds that act as separatrices for the 
Li-CN$\rightleftharpoons$Li-NC isomerizing reaction,
which is 
 a realistic molecular system.
We have shown how this tool adequately identifies the previous
manifolds as singularities when computed
for sufficiently long integration times.
Likewise,
we have demonstrated that, in general,
the adequate computation time must be large enough
compared to the inverse of the
characteristic stability exponent of the 
PO of interest.
Nonetheless, 
the previous 
criterion fails for the case
of bifurcating POs,
where the characteristic exponents
cancel
and then their inverses diverge.
Moreover,
two simplified models have been discussed
for a simplified description of the
results obtained with the original \emph{ab initio} PES.

First,
we have analyzed the performance of an
alternative PES
formed by Morse oscillators and also having 2 dof.
This equivalent model is also able to reproduce the same
structures that appear in one of the saddle-node bifurcations
of the system,
even when a constant moment of inertia for the angular coordinate
is considered.

As an additional simplification,
we considered a 1-dof model.
To start with,
we consider the potential energy
in this reduced dimensional model equal to that along the minimum
energy path.
Such a model is able to reproduce the manifolds
that emerge from the top of the energetic barrier of the system,
but not those that show up in another
saddle-node
bifurcation.

As the POs of interest are those with a
much faster radial motion than angular motion,
an adiabatic approach can be applied to 
the 2-dof Morse potential,
rendering a new energy surface for each quantized level.
This 1-dof model
is still able to successfully reproduce the main characteristics of
the invariant manifolds that emerge from the
top of the energetic barrier, as well as those
responsible for the appearance of the dynamical barrier
for energies~$E \ge E_{\rm SN_1}=
3440.6$\,cm$^{-1}$.
The shape of these last manifolds 
is 
strongly
influenced by the
number of excitations in the 
stretching motion associated with the
radial coordinate,~$n_R$.
Still, 
the Lagrangian descriptors are also equally able to
reproduce them.
In this case, 
the identification of the singularities in the plots of
the Lagrangian descriptors has enabled us to 
unveil the homoclinic intersection
that is responsible for the stable island
that is observed in the system with 2 dof.
Furthermore,
we have observed an imprint of the invariant manifolds at 
smaller energies~$E<E_{\rm SN_1}$,
which are responsible for 
the appearance of a local minimum
in the Lagrangian 
descriptors plots.
Similar results have been also previously
reported in Ref.~\onlinecite{GG20},
where the effect of the barrier height of a (unbounded) 
cubic potential is studied.
Note, nonetheless, that in our molecular system
the height of the energetic barrier cannot be arbitrarily tuned
as its value 
strongly depends on the vibrational energy and,
as a consequence,
is set by the adiabatic separation
of the different degrees of freedom.

To conclude,
let us remark that
the reduced dimensional models with 1 dof 
are 
not able to reproduce all the homoclinic and heteroclinic connections
that the system has in its full dimensionality.
This limitation is precisely responsible for the 
easy identification of the manifolds of interest.


\section*{Acknowledgements}\label{sec:thanks}
This work has been partially supported by the Spanish 
Ministry of Science, Innovation and Universities 
(Gobierno de Espa\~na) under Contract
No.~PGC2018-093854-BI00; 
by the People Program (Marie Curie Actions) of the European Union's 
Horizon 2020 Research and Innovation Program under Grant No.~734557;
and by the Comunidad de Madrid 
under Contract Grant No. APOYO-JOVENES-4L2UB6-53-29443N
(GeoCoSiM)
financed within the Plurianual Agreement with the
Universidad Polit\'ecnica de Madrid in the line
to improve the research of young doctors.
FB acknowledges financial support from the 
Ministerio de Ciencia, Innovaci\'on y Universidades
through the Severo Ochoa Programme for Centres of Excellence 
in R\&D (Grant No. CEX2019-000904-S). 
The authors also acknowledge computing resources at the
Magerit Supercomputer of the Universidad Polit\'ecnica de Madrid.

%
%
\appendix

\renewcommand{\figurename}{Figure A.\!\!}
\addto\captionsenglish{\renewcommand{\figurename}{Figure A.\!\!}}
\setcounter{figure}{0}
%
\section{Phase space for the two-dimensional system} \label{sec:2d}

\begin{figure}[hbtp!]
\includegraphics[width=0.85 \columnwidth, angle=0]{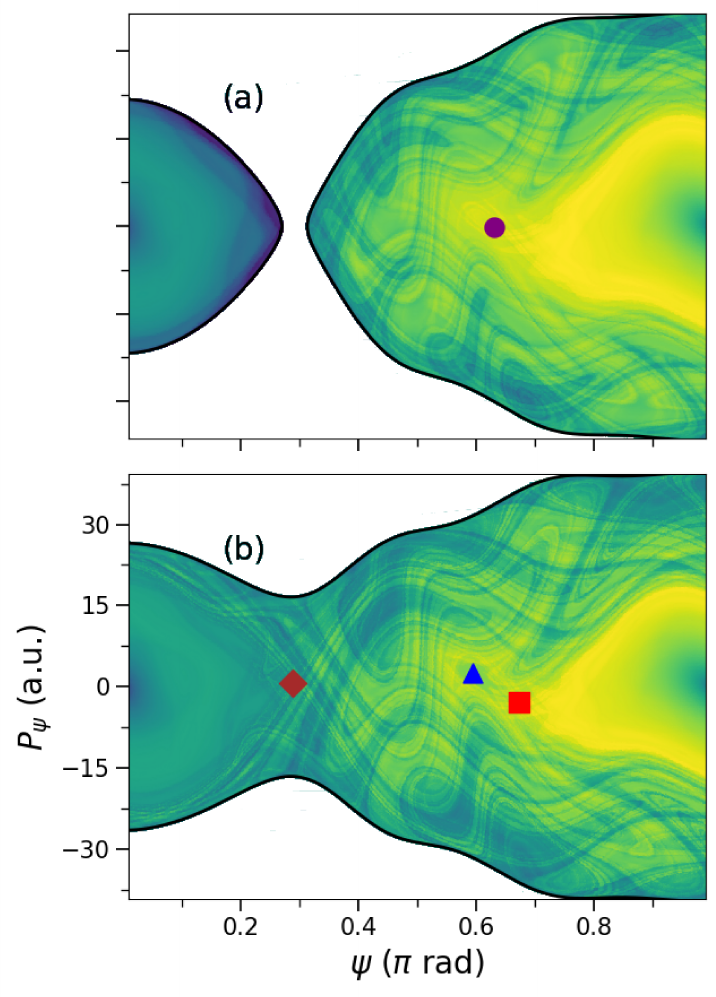}
\caption{Lagrangian descriptors given by Eq.~\eqref{eq:LD}
with~$p=0.4$ and~$\tau = 2 \times 10^4$\,a.u.
for (a)~$E=E_{\rm bif} = 3440.6$\,cm$^{-1}$and
for (b)~$E= 4000$\,cm$^{-1}$.
The purple circle, blue triangle, and red square 
show the position of the 
parabolic, elliptic and hyperbolic points associated with the three periodic orbits
related to the saddle-node bifurcation under study.
The brown diamond shows the position
of the unstable periodic orbit that is localized at the energetic barrier top 
and defines a recrossing-free dividing surface~\cite{Pechukas73, Pechukas77}.
%
}
\label{fig:LD_LiCN_2D_p04}
\end{figure}

In this appendix~\ref{sec:2d},
we discuss the Lagrangian descriptors (LDs)   
that are shown in Fig.~\ref{fig:2} 
in an extended region of the characteristic Poincar\'e surface of section
for the Li-CN$\rightleftharpoons$CN-Li isomerizing reaction.
For this purpose,
we show in 
Fig.~A.\ref{fig:LD_LiCN_2D_p04} the LDs
in the whole Poincar\'e surface of section
accessible at the corresponding energy. 
As can be seen, the phase space shown in
Fig.~A.\ref{fig:LD_LiCN_2D_p04}(a),
which is associated with the bifurcating energy~$E_{\rm bif}=3440.6$\,cm$^{-1}$,
is divided into two disconnected regions.
Consequently, no isomerization can take place,
as the 
Li atom does not have enough energy to overcome the
energetic barrier.
Notice also the different shape of the LDs
in the neighborhood of the
parabolic point (purple circle),
where the saddle-node bifurcation
(SNB) 
 discussed in detail in the 
main text
takes place.
When the SNB happens, 
a stability island appears
which opposes 
isomerization.

Contrarily, when the energy 
is larger than that of the main energetic barrier,
the phase space is formed by a single region,
which paves the way for isomerization;
this is the case
for~$E = 4000$\,cm$^{-1}$,
as shown in Fig.~A.\ref{fig:LD_LiCN_2D_p04}(b).
Notice that the structures that were shown in 
Fig.~\ref{fig:2} of the 
main text,
which emerge due to
the SNB,
cover quite densely the phase space of the system,
creating dozens of homoclinic and heteroclinic connections
while folding.

Let us remark on the presence of two important sets that 
are visible in Fig.~A.\ref{fig:LD_LiCN_2D_p04}(b).
On the one hand, 
the structures emanating from the SNB
can be clearly identified, namely
the 
stable island (where the elliptic point
shown as a blue triangle is embedded) 
and the invariant manifolds that surround it
(which emerge from the hyperbolic point shown as a red square)
can be clearly identified.
Consequently,
the influence of this structure on the system dynamics cannot be neglected.
On the other hand, 
The invariant manifolds associated with the saddle point
that lies at the top of the potential energy barrier
(brown diamond),
whose evolution is described in more detail in the main text,
can be also clearly identified.
Therefore, the model proposed with only 
1 dof 
is still capable of reproducing the geometry
that surrounds the 
two barriers that oppose the system reactivity:
the barrier top and  the SNB.

\renewcommand{\figurename}{Figure B.\!\!}
\addto\captionsenglish{\renewcommand{\figurename}{Figure B.\!\!}}
\setcounter{figure}{0}
\section{Phase space for the one-dimensional system
described with the effective potential} \label{sec:nR}

In this appendix~\ref{sec:nR}, 
we report on the phase space geometry 
associated with the 
1-dof
model for the LiCN molecule with the adiabatic potential
given by Eq.~\eqref{eq:Vmorse2}
for those vibrational numbers~$n_R$
that have been omitted in the main text,
namely, $n_R = 0, 1, 2, 4, 5, 7, 8$, and~9.
First,
we present in Sec.~\ref{subsec:1barrier}
the phase-space structure
for the single-barrier system,
which shows up when~$n_R = 0, 1,$ and~2.
Second,
Sec.~\ref{subsec:manybarriers}
is devoted to multiwell situations,
a situation that takes place
when~$n_R \ge~3$.
%
\subsection{Adiabatic potential with a single barrier}
\label{subsec:1barrier}

\begin{figure}[hbtp!]
\includegraphics[width=0.85 \columnwidth, angle=0]{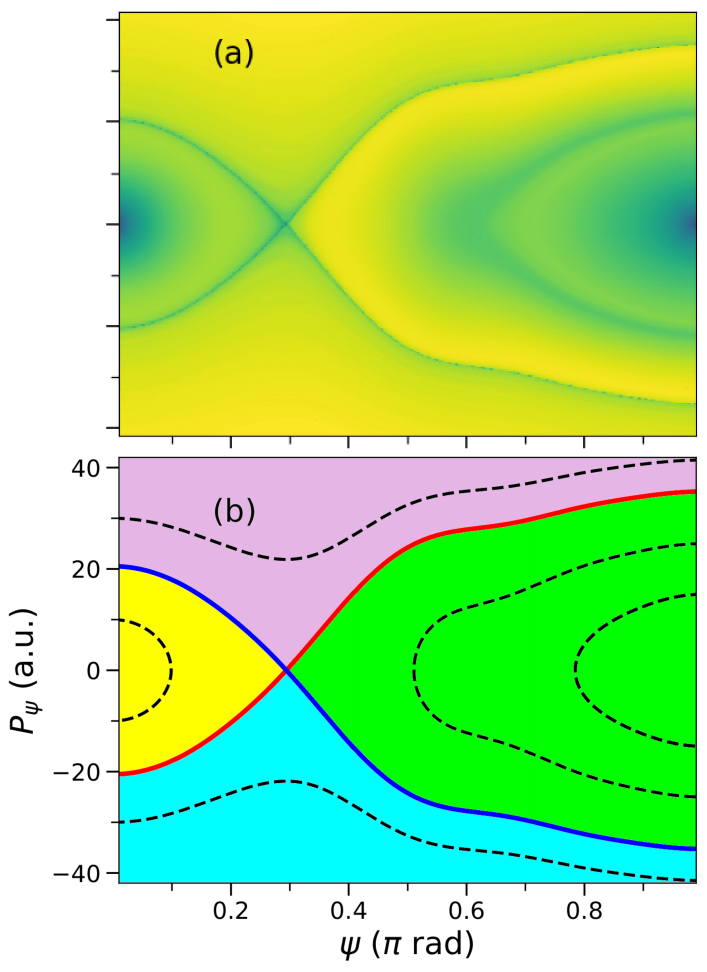}
\caption{
Phase space for the LiCN molecular system
described by the one-degree-of-freedom model
given by Eq.~\eqref{eq:Vmorse}  for
an effective adiabatic potential with $n_R=0$.
(a) Lagrangian descriptors computed for
$p=0.4$ and
$\tau = 2 \times 10^4$\,a.u..
(b) The stable (unstable) invariant manifolds that emanate
from the 
saddle 
point ($\psi \simeq 0.29 \pi$\,rad),
which are shown in blue (red),
partition the phase space
in the four colored regions.
}
\label{fig:nR0}
\end{figure}

Figures~B.\ref{fig:nR0}(a)-B.\ref{fig:nR2}(a)
show the value of the LDs
for~$n_R = 0, 1,$ and~2.
As can be seen,
the structure is very similar to that
already discussed in the Fig.~\ref{fig:mep}.
As in that case, 
the invariant manifolds show up as
singularities in the system.
These structures have been 
shown 
in red (unstable manifold) and blue (stable manifold)
continuous lines 
separately in the Figs.~B.\ref{fig:nR0}(b)-B.\ref{fig:nR2}(b),
along with some characteristic periodic orbits (dashed black lines).

\begin{figure}[hbtp!]
\includegraphics[width=0.85 \columnwidth, angle=0]{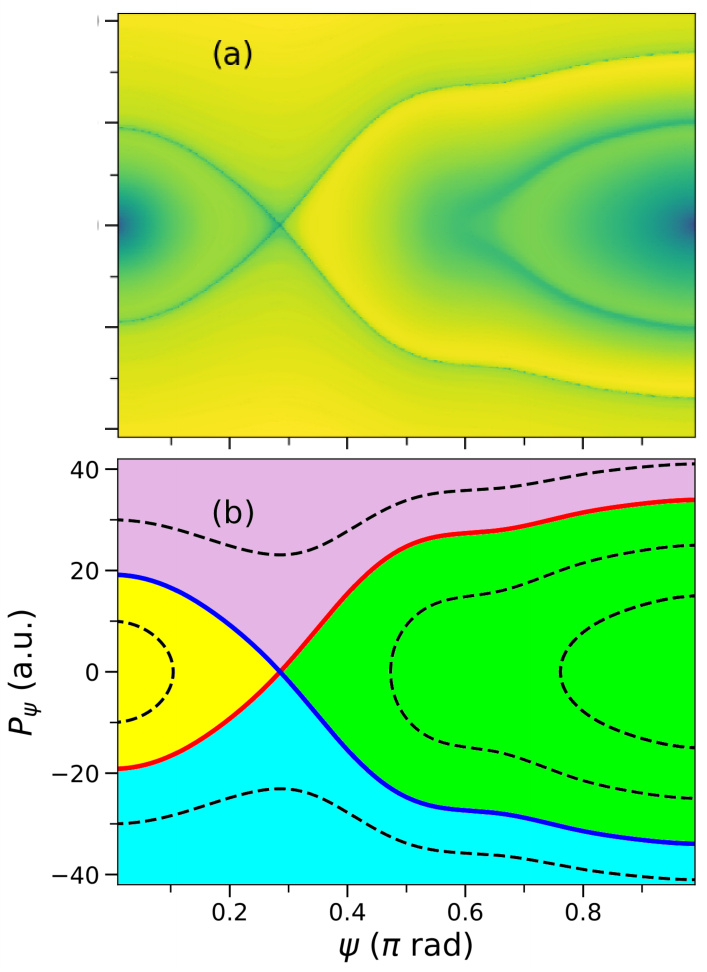}
\caption{
Same as Fig.~B.\ref{fig:nR0} but for~$n_R = 1$.
}
\label{fig:nR1}
\end{figure}

\begin{figure}[hbtp!]
\includegraphics[width=0.85 \columnwidth, angle=0]{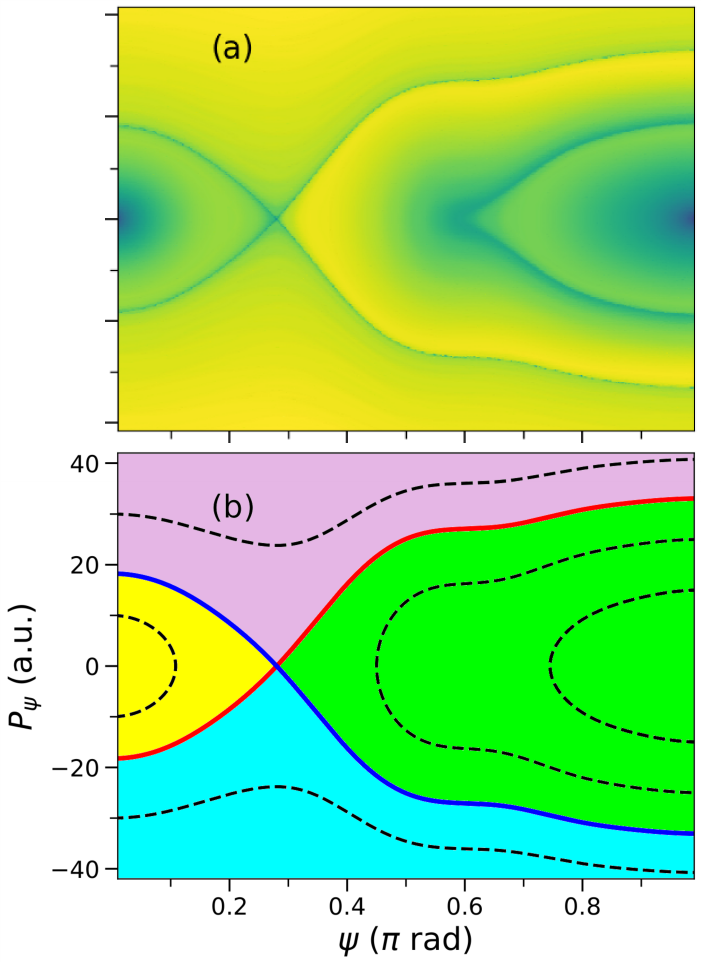}
\caption{
Same as Fig.~B.\ref{fig:nR0} but for~$n_R = 2$.
}
\label{fig:nR2}
\end{figure}

%
\subsection{Adiabatic potential with more than one energetic barrier}
\label{subsec:manybarriers}

\begin{figure}[hbtp!]
\includegraphics[width=0.85 \columnwidth, angle=0]{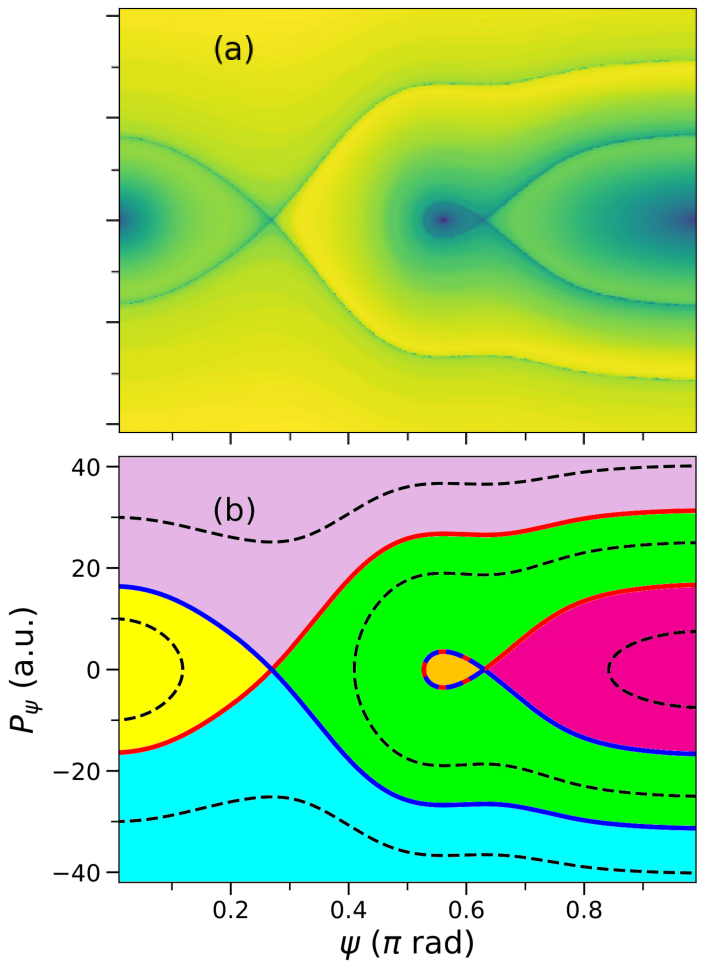}
\caption{
Same as Fig.~B.\ref{fig:nR0} for~$n_R = 4$.
Due to the saddle-node bifurcation, 
additional manifolds emerge from the
saddle point found at~$\psi \simeq 0.63 \pi$\,rad,
which further partition the phase space.
}
\label{fig:nR4}
\end{figure}

\begin{figure}[hbtp!]
\includegraphics[width=0.85 \columnwidth, angle=0]{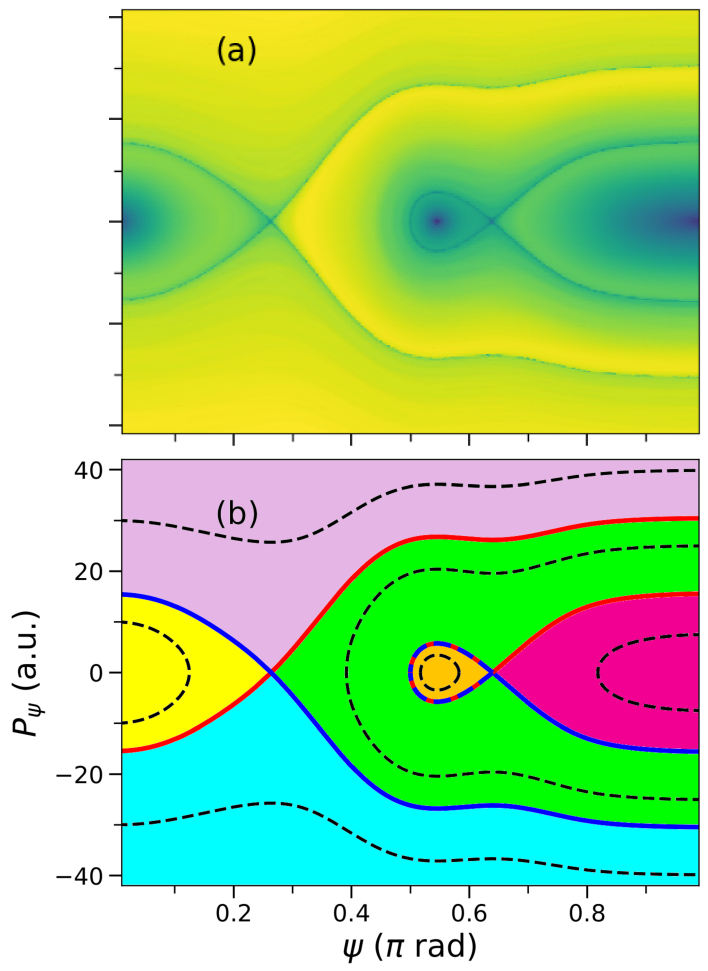}
\caption{
Same as Fig.~B.\ref{fig:nR4} for~$n_R = 5$.
}
\label{fig:nR5}
\end{figure}

Figures~B.\ref{fig:nR4}(a) and B.\ref{fig:nR5}(a) show the
LDs for~$n_R = 4$ and~5, respectively.
Contrary to the previous plots (cf. Figs.~B.\ref{fig:nR0}-B.\ref{fig:nR2}),
now two families of invariant manifolds are observed.
As already discussed in the main text,
the family of invariant manifolds
embedded in the green region is
associated with the saddle point that emerges
due to the SNB,
which introduces a secondary energetic barrier.
Notice that the area surrounded by the homoclinic
connection increases with~$n_R$,
as can be inferred by comparison with Figs.~\ref{fig:nR3}
and~\ref{fig:nR6},
where the results for~$n_R = 3$ and~6 are respectively shown.
%

\begin{figure}[hbtp!]
\includegraphics[width=0.85 \columnwidth, angle=0]{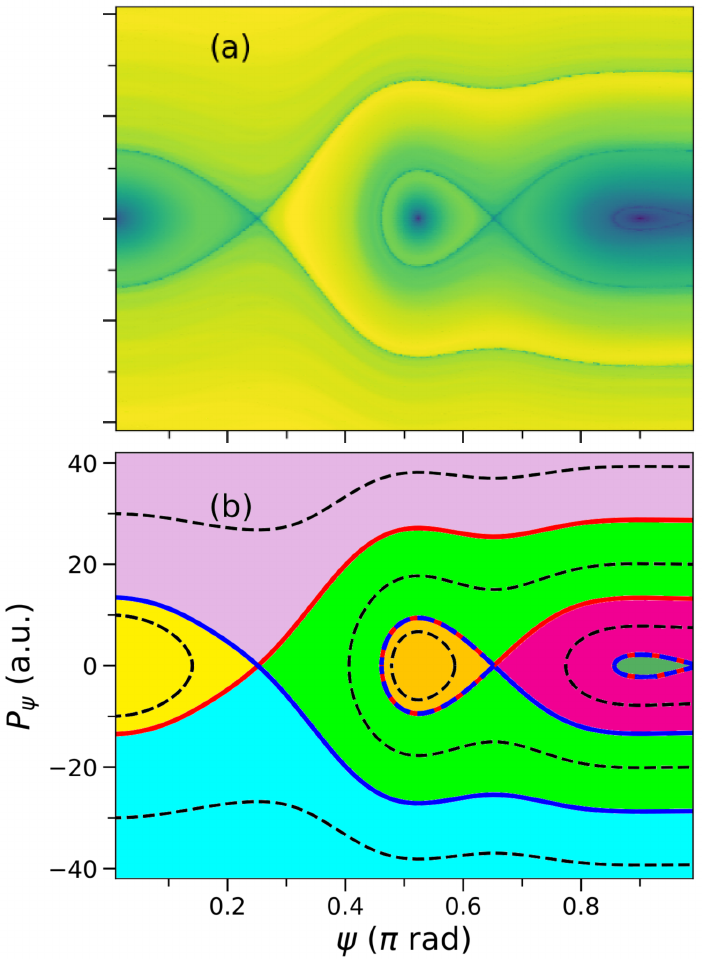}
\caption{
Same as Fig.~B.\ref{fig:nR4} for~$n_R = 7$.
The invariant manifolds associated with
the \emph{spurious} saddle-point that is found at~$\psi = \pi$\,rad,
and were absent for~$n_R \le 6$,
have been colored as dashed blue and red lines.
}
\label{fig:nR7}
\end{figure}

\begin{figure}[hbtp!]
\includegraphics[width=0.85 \columnwidth, angle=0]{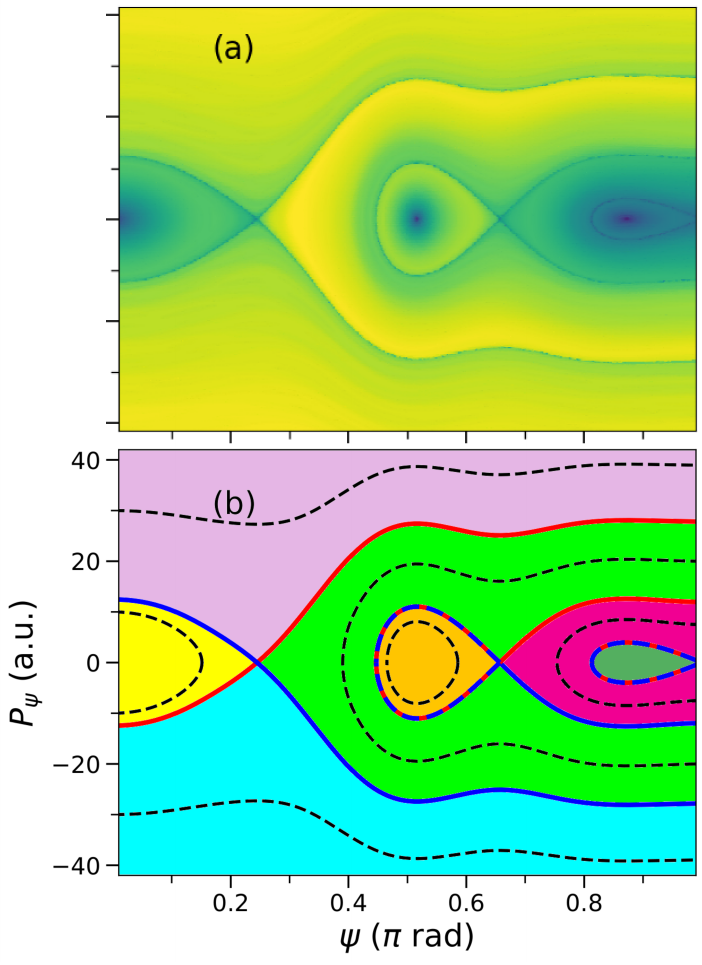}
\caption{
Same as Fig.~B.\ref{fig:nR7} but for~$n_R = 8$.
}
\label{fig:nR8}
\end{figure}

\begin{figure}[hbtp!]
\includegraphics[width=0.85 \columnwidth, angle=0]{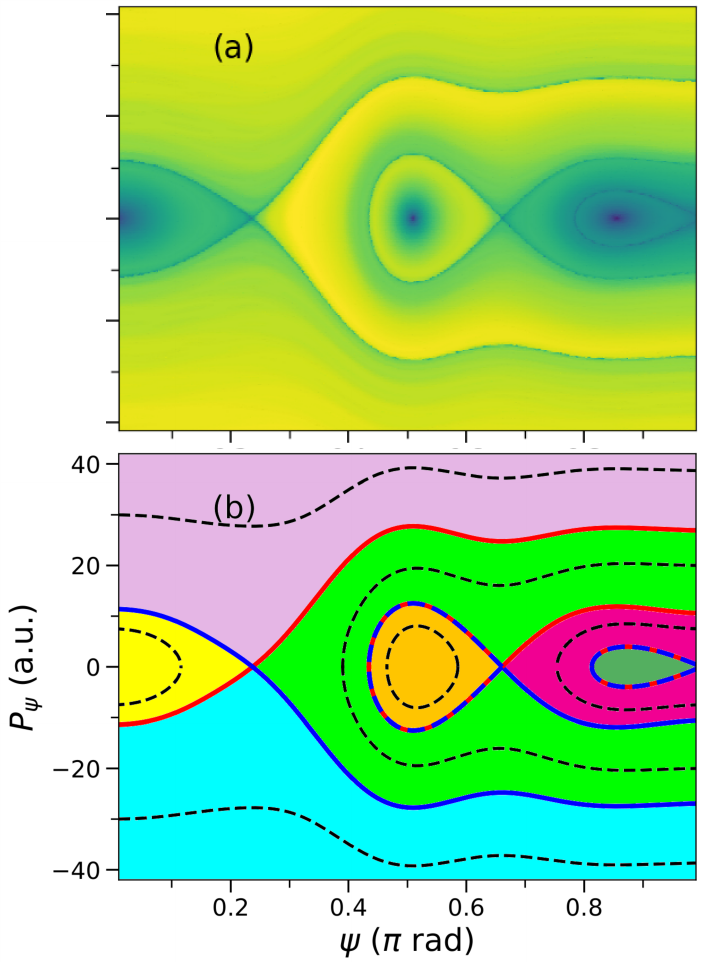}
\caption{
Same as Fig.~B.\ref{fig:nR7} but for~$n_R = 9$.
}
\label{fig:nR9}
\end{figure}

Figure~B.\ref{fig:nR7}(a) shows the LDs for~$n_R=7$.
As expected, the stability island 
has a larger area than in the previous cases discussed.
However, in this case an additional structure shows up
close to the well~$\vartheta = \pi$\,rad:
the well minimum moves from~$\vartheta = \pi$\,rad
to~$\vartheta \simeq 0.90\pi$\,rad, with
an additional tiny barrier appearing at~$\vartheta = \pi$\,rad,
where a third 
saddle 
point can be found.
As a consequence, an additional barrier emerges,
whose height  increases with~$n_R$.
The area occupied by the new stability region also
increases with~$n_R$,
as can be inferred from comparison with
 Figs.~B.\ref{fig:nR8} and~B.\ref{fig:nR9},
where the results for~$n_R = 8$ and~9 are shown.
Notice in Fig.~B.\ref{fig:nR7}(b),
where the invariant manifolds of Fig.~B.\ref{fig:nR7}(a)
are shown, that the second stability region is also
surrounded by a homoclinic orbit.
Contrary to the stability region discussed in connection
to the 
2-dof
system, the one that appears for~$n_R \ge 7$ 
does not  have an imprint on that system.
As a consequence, 
it is produced solely by the adiabatic approximation,
thus demonstrating the limitations of this approach.
%

%
\bibliography{ld_licn_1d}

\end{document}